\documentstyle[12pt,aasms4]{article}
\input{psfig.tex}  

\begin{document}

\title{DISTANCES, AGES AND EPOCH OF FORMATION OF GLOBULAR CLUSTERS
\footnote{Based on data from the Hipparcos astrometry satellite} }

\author{Eugenio Carretta and Raffaele G. Gratton}
\affil{Osservatorio Astronomico di Padova, Vicolo dell'Osservatorio 5, 35122
  Padova, ITALY\\
  carretta@pdmida.pd.astro.it and gratton@pdmida.pd.astro.it}
\authoremail{carretta@pdmida.pd.astro.it and gratton@pdmida.pd.astro.it}

\author{Gisella Clementini and Flavio Fusi Pecci
  \footnote{Also Stazione Astronomica, 09012 Capoterra, Cagliari, ITALY}}
\affil{Osservatorio Astronomico di Bologna, Via Ranzani 1, 
  40127 Bologna, ITALY\\
  gisella@astbo3.bo.astro.it and flavio@astbo3.bo.astro.it}
\authoremail{gisella@astbo3.bo.astro.it and flavio@astbo3.bo.astro.it}

\begin{abstract}

We review the results on distances and {\it absolute} ages of galactic
globular clusters (GCs) obtained after the release of the Hipparcos catalogue.
Several methods aimed at the definition of the Population II local distance
scale are discussed, and their results compared, exploiting new results for RR
Lyraes in the Large Magellanic Cloud (LMC). We find that the so-called {\it
Short} and {\it Long} Distance Scales may be reconciled whether a consistent
reddening scale is adopted for Cepheids and RR Lyrae variables in the LMC .

Emphasis is given in the paper to the discussion of distances and ages of GCs
derived using Hipparcos parallaxes of local subdwarfs. We find that the
selection criteria adopted to choose the local subdwarfs, as well as the size
of the corrections applied to existing systematic biases, are the main culprit
for the differences found among the various independent studies that first used
Hipparcos parallaxes and the subdwarf fitting technique. We also caution that
the absolute age of M92 (usually considered one of the oldest clusters) still
remains uncertain due to the lack of subdwarfs of comparable metallicity with
accurate parallaxes.

Distances and ages for the 9 clusters discussed in Paper I are re-derived
using an enlarged sample of local subdwarfs, which includes about 90$\%$ of
the metal-poor dwarfs with accurate parallaxes ($\Delta\pi/\pi \leq 0.12$) in
the whole Hipparcos catalogue. On average, our revised distance moduli are
decreased by 0.04 mag with respect to Paper I. The corresponding age of the
GCs is $t=11.5\pm 2.6$~Gyr, where the error bars refer to the 95\% confidence
range. The relation between zero age horizontal branch (ZAHB) absolute
magnitude and metallicity for the nine programme clusters turns out to be
$M_V(ZAHB) = (0.18\pm 0.09)({\rm [Fe/H]}+1.5) + (0.53\pm 0.12)$
Thanks to Hipparcos the major contribution to the total error budget
associated with the subdwarf fitting technique has been moved from parallaxes
to photometric calibrations, reddening and metallicity scale. This total
uncertainty still amounts to about $\pm 0.12$ mag.

We then compare the corresponding (true) LMC distance modulus $\mu_{\rm
LMC}=18.64\pm 0.12$\ mag with other existing determinations. We conclude that
at present the best estimate for the distance of the LMC is:
$$\mu_{\rm LMC}=18.54\pm 0.03\pm 0.06$$
suggesting that distances from the subdwarf fitting method are $\sim 1~\sigma$\
too long. Consequently, our best estimate for the age of the GCs is revised to:
$${\rm Age} = 12.9\pm 2.9~{\rm Gyr}$$
(95\% confidence range). The best relation between ZAHB absolute magnitude and
metallicity is:
$$M_V(ZAHB) = (0.18\pm 0.09)({\rm [Fe/H]}+1.5) + (0.63\pm 0.07)$$ 

Finally, we compare the ages of the GCs with the cosmic star formation rate
recently determined by studies of the Hubble Deep Field, exploiting the
determinations of $\Omega_M=0.3$\ and $\Omega_\Lambda=0.7$\ provided by type
Ia SNe surveys. We find that the epoch of formation of the GCs (at $z\sim 3$)
well matches the maximum of the star formation rate (SFR) for elliptical
galaxies in the HDF as determined by Franceschini et al. (1998).
\end{abstract}


\newpage

\section{INTRODUCTION}

In the last two years, new important sets of data have allowed for the first
time to connect the epoch of formation of local fossil remnants with evidences
from the remote Universe:
\begin{itemize}
\item The global history of star formation is now beginning to be
reconstructed, mainly thanks to data provided by the Canada-France Redshift
Survey (Lilly et al. 1996) and the Hubble Deep Field (HDF: Madau et al. 1996;
Madau, Pozzetti \& Dickinson 1998). Although quite large uncertainties still
exists, mainly related to the role of dust, these data robustly locate the
bulk of cosmic star formation at $z\geq 1$; for spheroidal systems, it is at
$z\sim 2-3$ (Franceschini et al. 1998).
\item Two independent, broad impact surveys of moderately high redshift type
Ia SNe (Schmidt et al. 1998; Pearlmutter et al. 1998) coupled to constraints
from Cosmic Microwave Background (Garnavich et al. 1998; Lasenby 1998) have
yielded very precise values for $\Omega _M\sim 0.3\pm 0.1$\ and $\Omega
_\Lambda\sim 0.7\pm 0.1$. Note that these small error bars are obtained
assuming a flat Universe. If confirmed, this very exciting result will allow a
much better definition of the Universe geometry than possible insofar
\item The ESA HIPPARCOS satellite has provided new, much more accurate values
for the trigonometric parallaxes of $\sim 118,000$\ nearby stars, including a
rather large number of Cepheids, horizontal branch stars, and subdwarfs. This
has allowed reliable estimates of distances and ages for the LMC and the
globular clusters (GC: Feast \& Catchpole 1997; Reid 1997, 1998; Gratton et
al. 1997a; Pont et al. 1998; Madore \& Freedman, 1998; Fernley et al. 1998a;
Chaboyer et al. 1998; Grundahl et al. 1998)
\item New data sets for variables in the LMC, from microlensing experiments
(MACHO, EROS, OGLE), as well as from other observing programs (Walker 1992;
Clementini et al. 2000) have allowed to clarify critical issues like the
r\^ole of reddening in distance derivations to this very important galaxy
\end{itemize}

The combined impact of these works is enormous, since first it allows to
reconcile the ages of GCs with the present and past expansion rate of the
Universe, and second opens new perspectives on galactic evolution studies. In
the present paper we re-examine the current status of knowledge about
distances and ages for GCs and discuss their impact on the cosmic distance
scale. Finally, we sketch the present evidences about the epoch of formation
of galactic GCs (assumed to be tracers of the earliest epoch of star formation
in our own Galaxy), comparing it with the current knowledge about the cosmic
star formation history.

\section{SHORT AND LONG DISTANCE SCALES AND THEIR IMPLICATIONS}

The status of the research on the absolute age of GCs, and its impact on the
choice of a cosmological model for the Universe, is summarized in Figure~1,
adapted and updated from Turner (1997).

As well known, the main problem in the derivation of ages for GCs is the
unambiguous definition of a reliable distance scale for population II stars
(Renzini, 1991), since a marked dichotomy exists insofar between a {\bf long
distance scale} and a {\bf short distance scale} (see VandenBerg, Bolte \&
Stetson 1996). Indeed, once the distance is fixed, absolute ages for the GCs
can be derived from the absolute magnitude of their turn off, $M_V(TO)$, which
is the evolutionary "clock".

Depending on which distance scale is adopted, a rather striking impact follows
on the cosmological framework. In fact, if the short distance scale is
adopted, the HB luminosity would be $M_V(HB) \sim 0.75$ at [Fe/H]$\simeq
-1.5$, GC absolute ages would be $\sim 16$ Gyr\footnote{For M92, the age
derived using the short distance scale would be as large as $\sim 18.5$ Gyr},
and a distance modulus of $\mu_{\rm LMC}\sim 18.25$ would be derived for the
LMC. This value is 0.25 mag smaller than that adopted
in extragalactic distance scales (e.g. the HST Key Project on Extragalactic
Cepheids), and implies a value for $H_0$ roughly in the range
65--85~km~s$^{-1}$Mpc$^{-1}$ (this residual range being mainly due to
ambiguities in the location of spiral galaxies within the Virgo cluster:
Tammann   1998). On the other side, if the long distance scale is adopted, the
HB luminosity would be $M_V(HB) \sim 0.5$ at [Fe/H]$\simeq -1.5$, GC's ages
would be $\sim 13$ Gyr, and the distance modulus to the LMC would be $\mu_{\rm
LMC}\sim 18.5$, consistent with $H_0$\ roughly in the range
55--75~km~s$^{-1}$Mpc$^{-1}$.

For this reason, Figure~1 is divided in two panels. In Panel {\it a} we
compared the age of GC derived before data from the Hipparcos mission became
available (Bolte \& Hogan 1995, Chaboyer et al. 1996; VandenBerg et al.
1996), with the age of a (flat) Universe provided by different values of
H$_0$, for different values of $\Omega_\Lambda = 1 - \Omega_M$, and within the
range indicated by the recent type Ia SNe data ($\Omega _M\sim 0.3\pm 0.1$).
The shaded region is the permitted area, obtained adopting values of
$H_0$\ consistent with the distance moduli used to derive the ages for the
GCs. While the rather large error bars indeed allow some region of overlap,
strong constraints are required for the epoch of formation of the GCs.

The advent of Hipparcos parallaxes has allowed to greatly improve one of the
most powerful (but up to now uncertain due to the lack of a proper data base)
method to derive distances: the GC subdwarf fitting technique. It will be
shown that once questions related to a correct handling of the data sample are
properly addressed, this method favours the "long distance scale", and the
derivation of younger ages for GCs. As shown in panel {\it b} of Figure~1, the
derived ages are now comfortably smaller than the age of the Universe.

While the distance scale provided by the subdwarf fitting method agrees with
the most accurate and up-to-now robust distance indicator: the Cepheids, it
must be emphasized that, so far, other recent results did not easily fit into
this reassuring and consistent picture. In fact, a number of distance scale
determinations still support the short distance scale. For instance, Hipparcos
proper motions for the field RR Lyraes have confirmed the results obtained by
ground-based statistical parallaxes. However, in the following we will show
that once reddenings for the RR Lyraes in the LMC are assumed consistently with
those used for the LMC Cepheids, this dichotomy is solved. This result is new
and will be discussed in some depth in the next Section.

\section{DISTANCES TO POPULATION II OBJECTS. A REVIEW}

Globular clusters, as well as most of the Population II objects, are too far
to allow a direct measure of their trigonometric parallaxes with the presently
available instrumental tools. Therefore, several {\it indirect} techniques
have been devised, in order to measure their distances. Some of them exploit
the classical {\it standard candles} existing among Pop. II objects (red giant
branch - RGB - stars, horizontal branch - HB - stars, RR Lyrae variables, main
sequence - MS - subdwarfs, white dwarfs - WD), others follow alternative
approaches. In this Section, we will consider several of these distance
indicators, both from population II and (for comparison) from population I
objects. We will postpone a thorough discussion of distances obtained via  main
sequence fitting to the next Section. Also, we will not consider distances
derived from calibration of luminosities entirely based on models (like e.g.
stellar or pulsational models). While most of the results we find here agree
with the model predictions, we prefer to keep results mainly based on
observations clearly separated from results entirely based on model
calculations. In this way, first, the results will not critically dependent on
the model assumptions, and, second, they may be used to test the model
predictions (see e.g. Gratton, 1998c; Castellani, 1999). On the other hand,
ages for clusters cannot be derived without reference to models: however, the
adopted calibrator - the MS turn-off luminosity - is a very robust theoretical
prediction.

The various distance indicators will be discussed through the comparison of
the values for the (true) distance modulus of the LMC, $\mu_{\rm LMC}$, they
lead to. When considering distances to the LMC derived from population II
objects the following two issues should be reminded :
\begin{itemize}
\item Distances to the LMC are finally founded on the RR Lyrae variables. The
absolute magnitude of RR Lyrae is known to depend on metallicity, but there is
no general agreement about the correct slope $dM_V(RR)/d$[Fe/H] of the
absolute magnitude-metallicity relation. In the following, we will assume a
value of $dM_V(RR)/d$[Fe/H]=$0.18\pm 0.09$, which is supported by most 
theoretical models (Chieffi, Straniero \& Limongi, 1998, Caloi, D'Antona \&
Mazzitelli 1997, Cassisi et al. 1997, VandenBerg, 1997), and Baade-Wesselink
absolute magnitudes (Fernley et al. 1998b), and it is only slightly larger
than derived from the GCs in M31 (Fusi Pecci et al. 1996). Both models and
observations suggests that the relation is probably steeper at metallicity
[Fe/H]$>-1$\ (however, this range is not relevant for the present discussion).
Anyway, the impact of 
this source of uncertainty ($< \pm 0.1$ mag/dex) on our discussion is small.   
In fact, our choice was to refer our absolute magnitude estimates to a   
metallicity of [Fe/H]$=-$1.5, that is a typical average value for most of the   
population II distance indicators: for instance, the average for our nine   
globular clusters is [Fe/H]$=-$1.42, that for HB stars with Hipparcos   
parallaxes considered by Gratton (1998b, see below) is $-$1.41, and that 
for the RR Lyrae used in the statistical parallaxes ranges from $-$1.53 to
$-$1.61, depending on the adopted sample. We argue that also RR Lyrae in the
bar of the LMC have on average a similar metallicity, albeit evidence is
admittedly not as strong as desired. However, on the whole we think that
uncertainties related  to these transformations are {\it on average} $< \pm$
0.02 mag, and cannot explain significant parts of the discrepancy between the
short and long distance scale, while of course they may be important for
individual clusters at the extremes of the metallicity range like M92 or 47
Tuc (in these cases, uncertainties in the distances may be as large as $\sim 
0.07$ mag, corresponding to $\sim 1$ Gyr)
\item Clementini et al. 1999 have recently derived an average magnitude of
$V=19.33\pm 0.02$\ for 75 RR Lyraes in two fields of the bar of the
LMC\footnote{This value is close to that estimated by the MACHO experiment
(Alcock et al. 1996), however, the band width used by MACHO is quite different
from the standard Johnson system, so that Alcock et al.'s color corrections
are more uncertain than in Clementini et al. photometry}. They compare this
value, which is assumed to correspond to an average metal abundance 
[Fe/H]=$-$1.5 dex (from the $\Delta$S analysis of a sample of double-mode
pulsators in the LMC), with Walker (1992) average dereddened magnitude of the
cluster RR Lyraes : $<V_0>=18.94\pm 0.04$\ mag (at an average metallicity of
[Fe/H]$=-1.9$\ and for an average reddening E(B$-$V) $\sim$0.09 mag). On the
assumption of a reddening value E(B$-$V)=0.10 mag for the LMC bar (Bessel,
1991) they conclude that (i) the two $<V_0>$ estimates are in very  good
agreement once the 0.4 dex difference in metal abundance is accounted for
(with a slope of 0.18 for the metallicity-luminosity relation for HB stars),
and that (ii) no evindence is found for a difference in luminosity between
field and cluster RR Lyraes in the LMC.  When combined with the faint absolute
magnitudes given by e.g. statistical parallaxes, these $<V_0>$ values for the
RR Lyrae's in LMC produce the so called {\it short distance scale}.

In a separate paper, (Clementini et al. 2000, in preparation), we present a
detailed description of this new LMC  RR Lyrae data set and discuss several
determinations of reddening towards the bar of the LMC. The (average) value
consistent with the analysis of the Cepheids (Caldwell \& Coulson 1985, Laney
\& Stobie 1993, Gieren, Fouqu\'e \& Gomez 1998) is $E(B-V)=0.07$; while a
slightly larger value of $E(B-V)=0.09$\ would give colors of the edges of the
instability strip consistent with those found for the LMC GCs (Walker 1992).
Finally, an even larger value : $E(B-V)=0.13$\ is given by the reddening maps
of Schlegel, Finkbeiner \& Davis (1998), Oestreicher, Gochermann \&
Schmidt-Kaler (1995) and Oestreicher \& Schmidt-Kaler (1996)\footnote{When
deriving reddening values from these maps, we added the foreground reddening
and half of the internal reddening, under the assumption that half of the
variables are in front of the absorbing layer, and half behind it.}. This
range in the reddening value obtained for the same locations within the LMC is
large, and may explain most of the scatter in the recent determinations of the
distance modulus of the LMC. In fact, the reddening value used by Walker
(1992), as well as Bessel's (1991), are on average $\sim$0.02-0.03 mag larger
than values consistent with the Cepheids analysis. Once corrected for this
effect, GC RR Lyraes are on average $0.03\pm 0.05$~mag brighter than the field
RR Lyraes, a difference entirely due to a single cluster (NGC 1841), that is
likely to be $\sim 0.2$~mag closer tu us than the LMC (Walker 1992).

Hereinafter, we will adopt the reddening scale given by Cepheids (0.07 mag)
which is also supported by a comparison of the colors of the edges of the
instability strip for the field LMC RR Lyraes with 
those defined by 
variables in the globular cluster M3 (Clementini et al. 2000).
 A major advantage of this choice is that it allows a consistent
comparison of distance scales from population I and II indicators; this is one
of the main tools of our analysis. With this reddening value {\bf we
obtain an average magnitude for RR Lyraes in the bar of the LMC of
$<V_0>=19.11\pm 0.02$}. The average metallicity of the bar RR Lyraes is not
well determined. In the following, we will assume a value of [Fe/H]$=-1.5\pm
0.2$, as the combination of Alcock et al. (1996, 1997) estimates, and of the 
$\Delta$S metal abundances for a sample of 
LMC double-mode pulsators (Bragaglia et al. 2000, in preparation).
\end{itemize}

\subsection{Population II distance indicators}

\noindent
a) {\it RR Lyrae and HB stars. Trigonometric Parallaxes}

\noindent
Gratton (1998b) found that the average absolute magnitude of a sample of 22
field metal-poor HB stars with trigonometric parallaxes measured by Hipparcos
is $M_V=+0.69\pm 0.10$ (at average metallicity [Fe/H]=$-1.41$). Popowski \&
Gould (1998b) reanalyzed Gratton's (1998b) HB sample, and after elimination of
all the red HB stars, because they may be contaminated by RGB stars, they
derive $M_V=+0.69\pm 0.15$ (at average metallicity [Fe/H]=$-1.62$). While the
revision by Popowski \& Gould may be questionable since the metallicity scale
for BHB stars is not well determined, Koen \& Laney (1998) have shown that the
distances derived by Gratton (1998b) are slightly underestimated because the
intrinsic scatter in HB magnitudes was neglected when correcting for the
Lutz-Kelker effect. The revised relation for RR Lyrae magnitude from
trigonometric parallaxes of HB stars is then:
$$M_V(RR)=0.18 ({\rm [Fe/H]}+1.5)+0.62\pm 0.11.$$ 
The corresponding distance modulus for the LMC is:
$$\mu_{\rm LMC}=18.49\pm 0.11$$

\noindent
b) {\it RR Lyrae and HB stars. Statistical Parallaxes}

\noindent
The Statistical Parallax method applied to galactic field RR Lyraes leads to a
faint zero-point of the RR Lyrae luminosity calibration. Using ground-based
proper motions, Layden et al. (1996) obtained $M_V=0.71 \pm 0.12$ at
[Fe/H]=$-$1.61. More recently, Popowski \& Gould (1998a) have re-analyzed
Layden et al.'s sample essentially confirming their results. The Hipparcos
based Statistical Parallax analysis by Fernley et al. (1998a) and Tsujimoto,
Miyamoto \& Yoshii (1997), also give a faint absolute magnitude. These
different estimates are well consistent to each other. However, since they use
very similar techniques and often the same observational data base, the error
bar of the average is essentially the same as that of the individual
determinations. A reasonable summary of these different results yields then
the following relation for RR Lyrae absolute magnitude:
$$M_V(RR)=0.18 ({\rm [Fe/H]}+1.5)+0.73\pm 0.12.$$
The corresponding distance modulus for the LMC is:
$$\mu_{\rm LMC}=18.38\pm 0.12$$

\noindent
c) {\it RR Lyrae and HB stars. The Baade-Wesselink method}.

\noindent
The Baade-Wesselink (B-W) method (Baade, 1926; Wesselink, 1969) uses the color,
light and radial velocity variations of an RR Lyrae variable during the
pulsation cycle, to derive its distance and absolute luminosity. This
technique in its two major variants, the surface brightness and the infrared
flux method (see Cacciari, Clementini \& Fernley 1992, and references therein),
has been applied to about 30 field's and to a few cluster variables (Liu \&
Janes 1990a,b, Jones et al. 1992, Cacciari et al. 1992, Skillen et al. 1993,
Storm, Carney \& Latham 1994), leading to a relatively mild slope of the
$M_V$,[Fe/H] relation, (as opposite to the steep slope of 0.30 found by
Sandage, 1993) and to a zero-point similar to that provided by statistical
parallaxes. Several attempts have been made to reconcile the B-W results with
the Cepheids distance scale. Fernley (1994) proposes a different value of the
conversion factor ({\it p}) between observed and true pulsation velocity, thus
getting a zero-point of the $M_V$,[Fe/H] relation 0.07 mag brighter than in
classical B-W analyses (see also Clementini et al. 1995). However, this only
accounts for about 1/3 of the original discrepancy. Feast (1997), using a
compilation of B-W literature data and adopting $M_V$ as independent variable,
derives $M_V = 0.37{\rm [Fe/H]} + 1.13$. When combined with Walker (1992) data
for the RR Lyraes in the LMC, this calibration provides $\mu_{\rm LMC}
=18.52$, in agreement with the classical modulus from Cepheids. However, Feast
procedure is rather questionable, since it attributes zero error to $M_V$,
which is by far the most uncertain quantity of the $M_V$,[Fe/H] relation.

A revision of the B-W results based on the assumption that optical and
near-infrared colors are better temperature indicators than the $V-K$ index,
is proposed by McNamara (1997). When applied to the RR Lyraes in the LMC, this
calibration gives $\mu_{\rm LMC} =18.54\pm 0.10$. However, systematic errors
are likely to affect McNamara procedure, since inconsistently derived
quantities are used.

Summarizing, the Baade-Wesselink method favours the following calibration
for the RR Lyrae magnitudes:
$$M_V(RR)=0.18 ({\rm [Fe/H]}+1.5)+0.71\pm 0.08.$$
(Clementini et al. 1995). The corresponding distance modulus for the LMC is:
$$\mu_{\rm LMC}=18.40\pm 0.08\pm 0.2,$$
where the first error bar is the internal one. However, given the theoretical
uncertainties related to the method, a more realistic error bar is perhaps 
$\sim 0.2$\ (Jones et al. 1992). This is the error we will adopt in our
discussion.

\noindent
d) {\it RR Lyrae and HB stars. Double-mode RR Lyraes}

\noindent
Thanks to the large amount of data from the MACHO experiment, Alcock et al.
(1997) were able to identify 73 double-mode RR Lyraes (RRd) near the bar of
the LMC. In these stars the ratio of the fundamental-to-first-overtone period
allows an accurate estimate of the star mass; they can thus be used to derive
pulsation-based luminosities of the HB. Alcock et al. assumed that their
variables are at Fundamental Blue Edge (FBE) of the instability strip and then
using the P$_{\rm FBE}$-[Fe/H] and log T$_{\rm eff}$-[Fe/H] relations by
Sandage (1993), obtain a calibration of the RR Lyrae luminosity with a bright
zero-point. This calibration leads to a distance modulus of the LMC of:
$$\mu_{\rm LMC}=18.48\pm 0.19,$$
for an assumed mean reddening of $E(B-V)=0.086$. This reddening is slightly
larger than the value used for Cepheids, so the comparison is somewhat
uncertain. However, since this source of uncertainty is much smaller than the
internal error bar of the determination, we may neglect it in our discussion.

\noindent
e) {\it Dynamical models for globular clusters}

\noindent
Astrometric distances to GCs can be derived from the comparison of proper
motion and radial velocity dispersion within each cluster, using King-Michie
type dynamical models (Rees, 1996). The method does not make use of {\it
standard candles} and is independent of stellar evolution models. However,
since results for individual clusters may be affected by large error bars, and
depend on cluster dynamical models (e.g. the incidence of binaries and the
presence of rotation), a large number of clusters should be analyzed in order
to increase the accuracy of the method. Rees (1996) derives distances based on
this technique for ten GCs. However, some of the analyzed clusters have rather
uncertain data. Chaboyer et al. (1998) restrict Rees' analysis to the 6
clusters with most reliable data, getting $M_V(RR)=0.59\pm 0.11$ at
[Fe/H]=$-$1.59.

Summarizing, the astrometric distances to GCs yields the following calibration
for the RR Lyrae magnitudes:
$$M_V(RR)=0.18 ({\rm [Fe/H]}+1.5)+0.61\pm 0.11.$$
The corresponding distance modulus for the LMC is:
$$\mu_{\rm LMC}=18.50\pm 0.11$$

\noindent
f) {\it The White Dwarfs Cooling Sequence}

\noindent
The distance  to a given cluster may be derived from the comparison of its WD
cooling sequence with a template sequence formed by local WDs with known
parallaxes and masses. This method has the advantage of being independent of
metallicity and age as well as of details of the convection theory. WDs have
been observed with HST in a number of GCs (M4, Richer et al. 1995; $\omega$
Cen, Elson, Gilmore \& Santiago, 1995; M15, De Marchi \& Paresce, 1995; NGC
6397, Cool, Piotto \& King, 1996; NGC 6752, Renzini et al. 1996; 47 Tuc,
Zoccali et al. 1998). So far, the method has been applied only to NGC 6752
(Renzini et al. 1996), and 47 Tuc (Zoccali et al. 1998).

The derived distances favour the short distance scale, and the following
calibration (not explicitly given in the original papers) can be obtained for
the RR Lyrae magnitudes:
$$M_V(RR)=0.18 ({\rm [Fe/H]}+1.5)+0.7\pm 0.15.$$
The corresponding distance modulus for the LMC is:
$$\mu_{\rm LMC}=18.4\pm 0.15$$

The major shortcomings of this technique are (i) the difficulties in comparing
HST colors for objects having very different apparent magnitude, and (ii) the
assumption that the calibrating WDs share the same mass as the GC WDs, an
assumption that may be criticized in view of the differences in the age of the
parent populations, and of the different techniques used to derive masses for
field and cluster WDs. These two effects are quite difficult to be estimated,
however our rather large (and subjective) error bars should be adequate to
represent them.

\subsection{Population I distance indicators}

Distances to the LMC can be also derived using population I distance
indicators. It is very important to compare these distances with those
provided by the population II objects.

\noindent
a) {\it Classical Cepheids. Trigonometric parallaxes}

\noindent
The Cepheid period-luminosity relation is traditionally the most accurate
method to derive distances to nearby galaxies. Feast \& Catchpole (1997) have
obtained a new P-L relation at V for classical Cepheids: $<M_V>=-2.81
\log P -1.43$, where the slope is that obtained by Caldwell \& Laney (1991)
from 88 Cepheids in the LMC, and the zero-point is based on the Hipparcos
parallaxes of a sample of Galactic Cepheids. When combined with an appropriate
correction for the metallicity dependence (+0.042 taken from Laney \& Stobie,
1994), this relation gives $\mu_{\rm LMC} =18.70\pm 0.10$. This result was
criticized by various authors (Szabados 1997; Madore \& Freedman 1998;
Oudmaijer, Groenewegen \& Schrijver 1998). However, Pont (1998) in
his thorough discussion of all these papers concluded that Feast \&
Catchpole analysis is unbiased (as demonstrated by appropriate MonteCarlo 
simulations), although Feast \&
Catchpole error bar is likely 
to be underestimated, a more appropriate value being $\pm 0.16$~mag. The same
conclusion has been reached analytically by Koen \& Laney (1998)

Summarizing, the Hipparcos based trigonometric parallaxes of Cepheids support
a distance modulus to the bar of the LMC of :
$$\mu_{\rm LMC}=18.70\pm 0.16$$

\noindent
b) {\it Classical Cepheids. Main Sequence Fitting}

\noindent
The zero point of the Period-Luminosity relation for Cepheids can be obtained
using Cepheids in clusters. Adopting this technique, Laney \& Stobie (1994)
obtained a distance modulus of $\mu_{\rm LMC}=18.49\pm 0.04 [\pm 0.04]$\,
where the former is the internal error, and the latter is the systematic one.
As pointed out by Pont (1998), this distance modulus corresponds to a Hyades
distance modulus of 3.27, slightly shorter than the very precise determination
by Hipparcos ($3.33\pm 0.01$). Once corrected for this effect, Laney \& Stobie
(1994) distance to LMC is modified to:
$$\mu_{\rm LMC}=18.55\pm 0.04 [\pm 0.04]$$

Possible uncertainties in this determination arise from Hipparcos revised
distance modulus for the Pleiades (Mermilliod et al. 1997), that, if adopted,
makes the cluster Cepheid P-L relation about 0.3 mag fainter. However,
Hipparcos distance modulus for the Pleiades might have an error bar much
larger than originally estimated, because all Pleiades stars lie in a small
region on the sky (Soderblom et al. 1998; Pinsonneault et al. 1998; Narayanan
\& Gould 1999); in this case individual parallax determinations cannot be
considered as independent, and the error bar should not decrease with the
square root of the number of stars. An error bar as large as $\pm 0.2$~mag may
likely be attributed to the Pleiades distance modulus from Hipparcos
parallaxes\footnote{According to Narayanan \& Gould, errors are much smaller
for the Hyades, due to the larger individual parallaxes, and their location on
the sky}.
 
\noindent
c) {\it Classical Cepheids. Baade-Wesselink Method}

\noindent
A further independent approach is followed by Gieren et al. (1998), who used
the infrared Barnes-Evans surface brightness technique to derive accurate
infrared distances to 34 Galactic Cepheids. These distances were then used to
determine period-luminosity relations at various passbands, which are compared
to analogous relations for Cepheids in the LMC. On the assumption of 
the same 
slope of the PL relation for the  Galactic and the LMC Cepheids, Gieren et
al. (1998) derive a distance modulus for the LMC of $\mu_{\rm LMC} =18.46\pm
0.06$. However, this distance modulus has been obtained with a slope of the
Period-Luminosity relation steeper than Caldwell \& Laney (1991) one. If
the latter is adopted, consistently with the other Cepheids distance
determinations, Gieren et al. distance modulus for the LMC becomes: $\mu_{\rm
LMC}=18.52\pm 0.06$. Using an empirical procedure based on the geometric
Baade-Wesselink method calibrated against high-precision data from
spectroscopic and interferometric techniques applied to Galactic Cepheids, Di
Benedetto (1997) obtained a corrected distance modulus for the LMC of $
\mu_{\rm LMC} = 18.58\pm 0.024,$ which agrees well with the (corrected) value
of Gieren et al. Note that Di Benedetto used a period-luminosity-color
($V-K$) relation for Cepheids. The slope of this relation cannot be easily
compared with the slope of the period-luminosity relation (at $V$) used by
Feast \& Catchpole. For this reason we prefer to apply no (uncertain)
corrections to the value given by Di Benedetto.

Averaging the result by Gieren et al. and Di Benedetto, we conclude that
the Baade-Wesselink method for Cepheids yields a distance modulus of: 
$$ \mu_{\rm LMC} = 18.55\pm 0.10,$$ 
for the LMC. The error bar has been (subjectively) increased to account for
systematic errors that may be rather large for this technique (note that this
is totally unrelevant in the following discussion).
 
\noindent
d) {\it SN1987A}

\noindent
The "light echo" of SN1987A has been used by several different authors to
derive an independent estimate of the LMC distance modulus. A rather small
value of $\mu_{\rm LMC} <18.37\pm 0.04$ (possibly increased to 18.44 for an
elliptical shape of the supernova expansion ring) was obtained by Gould
\& Uza (1998). This determination has been criticized by Panagia, Gilmozzi \&
Kirshner (1998), since Gould \& Uza used lines of different
excitation when comparing absolute and angular ring size. If lines of
comparable excitation are used, a larger distance modulus of $\mu_{\rm
LMC}=18.58 \pm 0.05$\ is derived (this value includes a small 0.03 mag
correction which takes into account the position of SN1987a within the LMC).
Panagia et al. (1998) distance modulus is compatible with the upper limit of
$\mu_{\rm LMC}<18.67$\ obtained by Lundqvist \& Sonneborn (1998).

The light echo of SN1987A is a very important method, because it is the only
one which is completely independent of reddening. The agreement between the
distance modulus derived from this method and the finally adopted one is then
a strong and crucial support to the Cepheid reddening scale used throughout the
present paper.

\noindent
e) {\it Eclipsing binaries}

\noindent
This is a very promising technique which uses the eclipsing binaries detected
in a number of systems (LMC, GCs, etc) to derive their distances (Paczy\'nski,
1996). In practice, the orbital parameters (period, inclination and
eccentricity), the luminosity-ratios, the size of the orbit, and the linear
radii of the two components are derived from the light and radial velocity
curves, and then combined with the surface brightness of each component
(inferred from the observed colors) to derive the distance. Since detached
eclipsing double line spectroscopic binaries have been discovered near the MS
turn-off of a number of GCs (Kaluzny et al. 1996, 1997), a direct measure of
the cluster distances via this technique should soon be feasible. The major
shortcoming of this method is how to properly derive the surface brightness of
the two components from the observed quantities (colors, line ratios etc.). So
far this technique has been applied to one eclipsing binary in the LMC
(HV2274, Guinan et al. 1998). The derived distance modulus is $\mu_{\rm
LMC}=18.30\pm 0.07$\, for an adopted reddening value of E(B$-$V)=0.12 mag (as
derived from the spectral distribution of the star). This reddening is only
slightly larger than derived from Cepheids in the same direction of HV2274
(0.106 $\pm$0.011 mag). Udalski et al. (1998b) have reanalized HV2274. They
derive $\mu_{\rm LMC}=18.22\pm 0.13$ assuming a reddening of 0.146 mag. Once
again, this underlines the large sensitivity to reddening of distance
determinations to the LMC.

These distance moduli, if confirmed once the other 10 binaries in Guinan et
al. sample will have been analyzed, favour the "short" distance scale.

\noindent
f) {\it The Red Clump}

\noindent
The red clump stars are the metal-rich counterpart of the HB stars. Red clump
stars have been detected in the Galactic Bulge by the OGLE microlensing
experiment (Udalski et al. 1993, Kiraga, Paczy\'nski \& Stanek, 1997). A well
defined red clump of solar-neighborhood stars is present in the CMD obtained
from Hipparcos data at an absolute magnitude of $M_V \sim 0.8$ mag (Jimenez,
Flynn \& Kotoneva, 1998). Since red clump stars are very numerous (the red
clump is the dominant post-main-sequence evolutionary phase for most stars),
and the observed dispersion in the mean clump magnitude is quite small, the red
clump can be used as a "standard candle" for distance estimates. The first to
use this technique for the LMC were Udalski et al. (1998a), who found a very
short distance modulus of $\mu_{\rm LMC}=18.08\pm 0.15$.

The applicability of the red clump method is based on two assumptions: first
it requires that the absolute magnitude of the red clump does not depend on
age and chemical composition and second, that the stellar populations in the
various systems do not significantly differ (in age and metallicity) from the
solar neighborhood red clump population. However, model calculations and
observations of open clusters show that quite large corrections are actually
required (Cole, 1998; Girardi et al. 1998). Furthermore, reddening is
important in this method too. Udalski himself (Udalski 1999) reconsidered this
issue, and obtained $\mu_{\rm LMC}=18.23\pm 0.05$\ using stars around four LMC
clusters having low reddening (reddenings used for these clusters are from the
maps by Schlegel et al. 1998, but in these directions the maps agree well with
the values given by Cepheids). On the other hand, the age and chemical
composition corrections used by Udalski are not universally accepted. Twarog,
Anthony-Twarog \& Bricker (1999) using metal-poor galactic open clusters with
distances determined via main sequence fitting obtain a much larger distance
modulus for the LMC ($\mu_{\rm LMC}=18.42\pm 0.16$). Romaniello et al. (1999)
used multi-band observations of the region around SN1987A, and carefully
discussed reddenings toward individual stars (reddening is rather large
towards this direction of the LMC). They found an even larger value of
$\mu_{\rm LMC}=18.59\pm 0.04\pm 0.08$. Finally, Piatti et al. (1999) suggested
that the RGB clump in the LMC has a vertical structure; if confirmed, this
would further complicate the data interpretation.

Given the large uncertainties still present in this method, we would give it
no weight in our discussion.

\noindent
g) {\it Miras}

\noindent
The period-luminosity relation for Miras based on Hipparcos parallaxes for a
sample of these variables, gives a distance modulus for the LMC of: $\mu_{\rm
LMC} =18.54\pm 0.18$ (Van Leeuwen et al. 1997), in good agreement with the
long distance scale.

\section{DISTANCES AND AGES OF GLOBULAR CLUSTERS USING HIPPARCOS SUBDWARFS}

\subsection{Previous results}

The subdwarf fitting technique is a very powerful method to derive distances
to GCs (Sandage 1970). However, until the release of the Hipparcos catalogue
this procedure was heavily hampered by the lack of local metal-poor dwarfs
with accurate parallaxes (VandenBerg et al. 1996). Hipparcos has definitely
enlarged the number and accuracy of the subdwarfs in the solar neighborhood,
which can be used in the subdwarf fitting. Thanks to Hipparcos, parallaxes are
no longer the major contribution to the total error of the subdwarf fitting
distances to GCs. In fact, besides parallaxes, a number of different
ingredients and assumptions enter into this technique, which all contribute to
the final result (i.e. the derived distance moduli), and to its accuracy. They
are: 1) magnitudes, colors, metallicities and reddenings of the calibrating
subdwarfs; 2) photometry, metallicity and reddening of the cluster stars; 3)
correction of the parallaxes for intervening observational biases; 4)
corrections for the contamination by undetected binaries in the subdwarf
sample and/or in the GC Main Sequences; and 5) evolutionary status of the
subdwarfs used in the fitting.

Three different groups had access to the Hipparcos database for the field
subdwarfs, before the whole catalogue was released to the general public: Reid
(1997; R97), Pont et al. 1998 (PMTV), and our team (Gratton et al. 1997a; Paper
I). After the publication of the Hipparcos catalogue, new analyses appeared
(Reid 1998: R98; Chaboyer et al. 1998: C98; Grundahl et al. 1998). A {\it
compendium} of the relevant parameters and of the main results achieved by
these studies is given in Table~\ref{t:tab1}. All the distance moduli are
apparent. If only true moduli were given in the original studies, apparent
ones were obtained using the quoted reddening (column 4 of Table~\ref{t:tab1})
and a value of 3.1 for the ratio of total-to-selective extinction. Also given
in Table~\ref{t:tab1} are the adopted metallicities, coming from the two major
abundance scales presently available for GCs, namely Zinn \& West (1984, as
subsequently updated and extended), and Carretta \& Gratton (1997, CG).
Other informations in Table~\ref{t:tab1} concern the number of calibrating
subdwarfs used in the fitting, together with the spanned metallicity range,
the apparent moduli corrected to account for the possible presence of
contaminating binaries in the subdwarfs sample, and references for both
cluster mean ridge lines and distance moduli.

Table~\ref{t:tab1} allows a quick comparison of the results obtained by
different authors. PMTV (1 cluster, M92) essentially confirm the distance and
age estimates based on the ground-based parallax observations; Paper I, R97,
R98, C98, and Grundahl et al. (1998) consistently derive $longer$ distances
and, in turn, $younger$ ages for the clusters in their samples. In the
following we briefly discuss and compare these independent studies.

{\bf (a) Reid 1997}: R97 primary calibrating subdwarfs include 15 stars with
parallaxes measured by Hipparcos with $\Delta\pi/\pi=0.12$, plus 3 stars from
the Yale General Catalogue of Trigonometric Parallaxes (van Altena,
Truen-liang Lee \& Hoffleit 1995), at same precision limit. Photometric data
and low dispersion spectroscopic metallicities for the calibrating subdwarfs
were taken from Carney et al. (1994). Metal abundances for the 7 studied GCs
were instead from Zinn \& West (1984), and mainly based on low dispersion
indexes. A direct comparison of the two metallicity scales is not possible,
but they are certainly not the same, which is one of the major drawbacks of
this first analysis. Appropriate Lutz-Kelker corrections were included. Only
metal-poor and intermediate-metallicity clusters are studied, with reddening
estimates from a compilation of different literature sources. No attempt was
made to select clusters at low reddening contamination. All subdwarfs are used
in the fitting, independently of their evolutionary status (4 of R97 primary
calibrators are evolved objects with $M_V < 5.0$), or their binary status (but
binaries are excluded from the fittings).

{\bf (b) Gratton et al. (1997a; Paper I)} derived subdwarf fitting distances
for 9 clusters, six of which are in common with R97. There are a number of
differences and, we think, improvements in Paper I with respect to R97
analysis:
\begin{itemize}
\item a larger sample of subdwarfs (34 stars, 22 of them with [Fe/H]$<-0.9$)
\item a careful reduction of the photometric data for the field subdwarfs to a 
consistent scale
\item the strict homogeneity in the metal abundances of field $and$ cluster
stars (see however Section 4.8). High resolution spectra were obtained for
most of the subdwarfs in their sample, and accurate abundances of Fe, O and
$\alpha$-elements were derived (Clementini et al. 1999, CGCS99), following a
procedure totally consistent to that used by CG for giants in GCs. For the
remaining objects, new metallicities on the same metallicity scale were
obtained through the re-calibration of low resolution abundance indicators
(see CGCS99)
\item a careful revision of the cluster photometries in order to eliminate
possible inconsistencies. Nevertheless, an unambiguous and reliable
photometric zero-point still lacks for some clusters (e.g. M92, see below
Section 4.6)
\item only clusters with interstellar reddening not exceeding 0.05 mag were
analysed. Furthermore, in order to reduce reddening uncertainties still
existing even for the best studied clusters, new reddening estimates were
derived from the Str\"omgren and $H_\beta$\ photometry of B-F stars projected
on the sky within 2~degrees from each cluster, and averaged to the previous
literature estimates
\item a careful procedure was devised in order to detect unknown binaries.
Corrections were thus calculated as the product of the probability for a star
to be a binary, and of the average correction for each binary, derived from
the offset of the known binaries from the $(B-V)_{Mv=6}-$[Fe/H] relation
\item finally, only {\it bona-fide} single stars on the unevolved portion of
the main sequence ($M_V < 5.5$) were used in the fittings
\end{itemize}
On average there is a reasonable agreement between Paper I distance moduli and
R97, the mean difference being $-0.08 \pm 0.04$ mag (six clusters, rms scatter
of 0.09 mag): the distance scale of Paper I is then shorter than that of R97

{\bf (c) Reid (1998)} partially revises R97 analysis and extends it to the
metal-rich clusters, applying the MS fitting technique to 7 clusters, 3 of
which were not included in R97. Besides an enlarged sample of calibrating
subdwarfs and the improved treatment of the Lutz-Kelker corrections, the most
relevant change in R98 analysis is the adoption of a metal abundance scale
directly tied to high resolution spectroscopic analyses (i.e. CG) 
for the GCs, and of a very similar one for the subdwarf sample.
Taking into account the existing differences in the calibrating sample, the
adopted reddening and bias corrections, and within the quoted uncertainties,
there is an overall agreement between R98 and Paper I (see Table~\ref{t:tab1}
and R98 for further details). We note that R98 adoption, of a consistent
metallicity scale for clusters and field subdwarfs, leads to halve the
difference between R98 and Paper I distance moduli of NGC 6205, 6752 and 5904.

{\bf (d) Pont et al. (1998)} had access to the whole Hipparcos catalogue just
after the final data reduction. Their calibrating subdwarfs were extracted
from the "merger" of two distinct lists. The first one (330 stars) was formed
by subdwarf candidates selected from various sources before the Hipparcos
parallaxes were known; the second one (216 stars) was selected {\it a
posteriori} from the whole Hipparcos catalogue on the basis of parallaxes and
colors. Among these more than 500 candidates, PMTV selected 17 stars with
claimed metallicity [Fe/H]$< -1.8$, that they used to define a fiducial
sequence (from the unevolved region below the turn-off point up to the
subgiant phase) to match the mean locus of the GC M92. Their result was a
distance modulus (14.68 $\pm$ 0.08) very similar to that obtained by previous
studies based on subdwarf parallaxes measured from the ground, and an age of
14 Gyrs. Differences in the subdwarfs calibrating samples are not the origin
of this contrasting result which are mainly due to PMTV different, and
sometimes arbitrary assumptions on (i) whether and how large bias corrections
should be applied to the Hipparcos parallaxes (this conversely reflects
differences on how the calibrating subdwarfs were originally selected), (ii)
which corrections should be applied to account for undetected binaries
contaminating the subdwarf sample, and (iii) whether only unevolved subdwarfs
should be used or also subgiant stars should be included in the fittings
(Gratton et al. 1997b, Gratton, Carretta \& Clementini, 1999).

{\bf (d) Chaboyer et al. (1998)} initially selected stars using an {\it a
priori} sample drawn from the whole Hipparcos catalogue. They then restricted
their analysis to the unevolved stars with spectroscopic abundances, and
discarded all known or suspected binaries (but they prefered to apply no
correction for undetected binaries). With this procedure they end up with a
rather small sample of only 7 subdwatfs, all in a quite restricted metallicity
range. For this reason they only give distances to three GCs (NGC6752, M5 and
M13) having a metal abundance within the range covered by the observed
subdwarfs. Rather than using homogenous abundance and reddening analyses for
field stars and GCs, they prefer to discuss each star individually, taking 
what they consider to be the best guess for each star. While some
inhomogeneity may indeed be introduced by this practice, they try to account
for it with the adoption of  error bars somewhat larger than those assumed  by
other authors. On the whole, their results agree very well with those of Paper
I and of the present paper, although their distance modulus for M5 is slightly
shorter than ours.

{\bf (e) Grundahl et al. 1998}: much less details are available for the
analysis of Grundahl et al. (1998), who derived a distance to M13 of
$(m-M)_V=14.44\pm 0.10$, in good agreement with Paper I, this paper, and C98.

Concluding this discussion, we wish point out two major issues that have been
poorly stressed insofar:
\begin{itemize}
\item all these distance estimates are in agreement with or even longer than
the long distance scale ($0.19<M_V(HB)<0.51$\ at [Fe/H]$=-1.5$\ and with a
slope of 0.18 for the metallicity-luminosity relation for HB stars). This is
much brighter than given by the short distance scale ($M_V(HB) \sim 0.75$ at
[Fe/H]$=-1.5$). In turn, the ages implied for M92 range from 11.4 to 15.4~Gyr
on the scale used throughout this paper, while the age of M92 given by the
short distance scale is $\sim 18.5$~Gyr
\item {\bf Hipparcos parallaxes are systematically smaller than those
previously obtained from ground}. This, by itself, directly translates into a
"stretching'' of the GC distances, and, in turn, in a 2-3 Gyrs decrease of
their ages (in Paper I we computed a mean magnitude difference of $0.20\pm
0.04$~mag, limiting our attention to stars with $\Delta \pi/\pi<0.12$). Large,
corrections must be applied to transform these smaller parallaxes into
distance moduli that are very similar to the older values obtained from larger
parallaxes
\end{itemize}

\subsection{New results using the overall HIPPARCOS database} 

Since the whole Hipparcos catalogue is now a public resource, it is possible
to exploit its entire database in order to improve the subdwarf fitting
analysis. Two different routes may be followed: i) to use an {\it a priori}
sample where stars are selected according to criteria not based on Hipparcos
parallaxes (as in R97, R98 and Paper I), ii) to use an {\it a posteriori}
sample extracted from the whole Hipparcos color magnitude diagram (as in C98,
and, in part, in PMTV).

The main difference between these two approaches is the number and relevance
of the statistical biases associated to intervening selection effects. In
fact, while in a {\it a priori} selected sample, selection criteria (and
corresponding biases) are known as they are defined by the investigators
themselves, any {\it a posteriori} selected Hipparcos sample is hampered by
the poor knowledge of the selection criteria originally applied to form
the Hipparcos catalogue itself, and by the increasing incompleteness of the
catalogue towards faint magnitudes ($V>7$). In particular, the metallicity of
an {\it a posteriori} selected sample of subdwarfs, as well as the rate and
distribution of secondary components of the contaminating binaries, are rather
poorly defined, thus rendering very uncertain the correction of the
corresponding systematic effects.

In the following Sections, we discuss these two approaches in details.
According to the procedure outlined in Paper I, preference is given to the
{\it a priori} sample approach, while the {\it a posteriori} sample is used to
show how uncertainties, or different assumption about reddening, metallicity
scale and binary corrections, can affect the analysis, and to test the
completeness achieved by the {\it a priori} selected sample.

\subsubsection{The {\it a priori} sample}

Stars of the {\it a priori} sample were extracted from the catalogues of
Gratton, Carretta \& Castelli (1997, GCC), Carney et al. (1994), Axer,
Fuhrmann \& Gehren (1994), Ryan \& Norris (1991) and Schuster \& Nissen
(1989). We selected all the stars  with $V < 12$ and a
metallicity [Fe/H]$< -0.9$ (after correcting the original values to GCC
scale). This sample includes about 400 stars and constitutes almost a proper
motion limited sample. About 60$\%$ of the sample (246 objects) has parallaxes
measured by Hipparcos. Most of the missing stars are faint objects. In fact,
all stars with $V<9$ and about 88$\%$ of stars with $9 < V < 10$ in our
original sample are contained in the Hipparcos catalogue. This fraction
decreases to about 50$\%$ and less than 30$\%$ in the magnitude ranges $10 < V
< 11$ and $11 < V < 12$, respectively.

Thus, as far as the metal-poor dwarfs are concerned, the Hipparcos catalogue
seems to be almost complete, or at least adequately representative, down to
$V<10$. The introduction of a selection criterion based on the parallax
accuracy changes dramatically the situation, because the median errors of
Hipparcos parallaxes increase with apparent magnitude (they are less than 1
mas for $V\leq 7.5$, around 1.3 mas $V\sim 9$, and $>2$ mas for $V>10.5$). If,
following Paper I, we cut the sample to stars with $\Delta \pi/\pi<0.12$
\footnote{While this cut is arbitrary, we found it to be a good compromise
to achieve a large enough sample of mainly metal-poor objects, with still small
individual errors and Lutz-Kelker corrections. The extension 
to objects with $\Delta \pi/\pi<0.2$\ does not provide additional useful
informations}, we are left with only 56 objects, all brighter than $V=10.5$.
Stars fainter than this limit are of only limited use for the present purpose.

In their fitting of the M92 sequence PMTV used also a few subgiant stars with
$M_V<3.6$. Seven subgiants are present in our {\it a priori} sample. Since
stars with $M_V<3.6$ must be brighter than $V \leq 8.8$ in order to have
parallaxes with $\Delta \pi/\pi<0.12$, and since our sample seems quite
complete for V$<9$ (see next subsection), we are confident that
most of the metal-poor subgiant stars with good parallax are present in our
sample. Unfortunately, a good parallax is not a sufficient condition to use
these stars for distance determinations. In fact:
\begin{itemize}
\item HD17072 is probably a HB star (see Gratton 1998b, Carney, Lee \& Habgood
1998); HD6755 is too evolved, and HD89499 is out of any possible fitting
relation
\item reddening estimates for HD132475, HD140283, HD189558 (all included in
the PMTV sample) and HD211998 are quite large and uncertain
\item HD211998 is a very close visual binary
\end{itemize}

More interesting for the purpose of deriving GC absolute ages are the
unevolved stars close to the Zero Age MS ($5<M_V<8$). We have 30 such objects
with $\Delta \pi/\pi<0.12$ in our {\it a priori} sample. None of them has
metal abundance [Fe/H]$< -2$, and 9 (2 of which are binaries) have [Fe/H]$<
-1.5$. In Paper I we considered 33 stars in this absolute magnitude range;
however, 12 of them had  metallicity [Fe/H]$>-0.9$ and are then excluded from
the present discussion. Seven of these stars (2 binaries) have [Fe/H]$< -1.5$.
We have then 9 additional metal-poor stars with good parallaxes; 2 of them are
{\it bona fide} single stars with [Fe/H]$< -1.5$.

Finally, we note that all 9 stars with [Fe/H]$<-1.5$ have $(B-V)_{Mv=6}<0.65$.
On the other side, approximately half of the 21 stars with
$-1.5<$[Fe/H]$<-0.9$ have $(B-V)_{Mv=6} < 0.65$ (11 stars, 2 binaries) and the
remaining  half (10 stars, 4 of which are binaries) have $(B-V)_{Mv=6} >
0.65$. This suggests that, in this range of magnitude, selecting stars with
$(B-V)_{Mv=6}<0.65$ corresponds to obtaining a complete sample of objects with
[Fe/H]$<-1.5$. However, samples selected using these criteria should be
considerably contaminated by more metal-rich stars.

Table~\ref{t:tab2} summarizes the relevant data for the 56 stars with $\Delta
\pi/\pi<0.12$ in the {\it a priori} sample.

$V$ magnitudes are the average of Carney et al. (1994), Ryan \& Norris (1991)
and Schuster \& Nissen (1989) values, together with values from both Hipparcos
and Tycho catalogues. Reddening values are obtained by averaging estimates
from the 3 above-mentioned sources, or by adopting a cosecant law (Bond 1980),
when such estimates are lacking. Absolute magnitudes include Lutz-Kelker
corrections (see Sections 4.3 and 4.4.1). $B-V$ colors are the weighted average
of Carney et al., Ryan \& Norris (with an arbitrarily adopted error of 0.01
mag) and Hipparcos/Tycho values (with the errors quoted in the catalogue).
Colors in Table~\ref{t:tab2} are de-reddened and errors include an adopted
uncertainty due to reddening of 0.015 mag, for all stars. Whenever available,
metallicities were from high resolution spectroscopic analyses (e.g. GCC, Axer
et al., 1994, etc.), and tied to GCC/CG homogeneous metallicity scale. A
weighted mean of the metallicities from Carney et al. (cross correlation
dips), Ryan \& Norris (low dispersion spectroscopy) and Schuster \& Nissen
(Str\"omgren photometry), corrected to GCC scale, was adopted for the stars
lacking high resolution spectroscopy. An uncertainty of 0.15 dex, was assumed
in this case. Finally, information on binarity is given, with the same meaning
of Table~1 of Paper I.

\subsubsection{The {\it a posteriori} sample}

As an alternative approach, a sample of metal-poor stars was extracted from
the total color-magnitude diagram of the Hipparcos catalogue. This {\it a
posteriori} approach is similar to the procedure used by PMTV to form their
second sub-sample of 216 stars, and to the procedure followed by C98. From the
available database we first extracted all stars with positive parallax and
$\Delta \pi/\pi<0.12$ in the magnitude range $5.0<M_V<8.0$\ mag (4342 stars;
this magnitude range should include mainly unevolved MS stars). Metal-poor
candidates were then identified by inferring the metallicity from the B$-$V
color given in the catalogue.

By shifting the location of an unevolved star in the color-magnitude diagram
parallel to the main sequence (see Section 3 of Paper I), it is possible to
derive its $B-V$\ at an absolute magnitude of $M_V=6$, $(B-V)_{Mv=6}$. In the
range of interest, this color is predicted by theoretical models to be an
approximately quadratic function of metallicity, decreasing the metal abundance
giving bluer MSs (see equation 4 in Paper I). This relation was used to
extract the {\it metal-poor} subdwarf candidates from the whole {\it a
posteriori} selected unevolved stars with $\Delta \pi/\pi<0.12$. We found that
the number of metal-poor stars below the MS is very small: there are only 52
stars (1.2$\%$) with (B$-$V)$_{Mv=6} \leq 0.65$ (i.e. [Fe/H]$<-1$) and $\Delta
\pi/\pi<0.12$. We caution, however, that metallicities deduced from colors may
be overestimated (and, correspondingly, the number of metal-poor stars
underestimated) since stars can be made redder by the interstellar reddening
or by the presence of binary companions. These effects are balanced only in
part by the Lutz-Kelker bias (stars that for erroneous measurements are below
their true position have higher probability of being selected thus making the
derived sequence bluer).

The problem is further complicated since the metallicity distribution of the
stars in the solar neighbourhood is strongly skewed toward solar values. As a
consequence, symmetrically distributed random color errors shift more stars in
the lower metallicity bin than in the high metallicity one. An additional
complication is the non-gaussian distribution of errors in the color data
sets; e.g. in the Tycho catalogue there is a small but not negligible number
of stars ($\sim 1$\%) with color errors that may be as large as several tenths
of a magnitude. Given these large errors, solar metallicity stars may be
sometime erroneously interpreted as extremely metal-poor objects. The fraction
of stars with large random color errors [$\Delta (B-V)>0.05$ mag] in the Tycho
catalogue is comparable to (actually larger than) the fraction of metal-poor
stars (with [Fe/H]$<-1$) in the Hipparcos catalogue. A large contamination of
spurious objects is then expected for the bluest bins in $(B-V)_{Mv=6}$.

This {\it metallicity bias} has a large impact on the {\it a posteriori}
sample, where metallicities are derived from colors, while does not affect
stars in the {\it a priori} sample, whose metallicities are known
independently of colors. This is an important difference and a reason to
definitely prefer the {\it a priori} approach.

\subsection{Detecting, understanding and correcting for biases}

{\it A priori} and {\it a posteriori} samples are both affected by a number of
biases that must be taken into account in order to derive statistically
corrected distances (and ages) of GCs. They are briefly discussed below.
\par\noindent
{\it 1. Lutz-Kelker effect and Malmquist bias}\\
Average luminosities for sample of stars selected according to their absolute
magnitudes (i.e. parallaxes) are affected by the Lutz-Kelker (1973) effect.
This bias arises from the combination of measurement errors in the parallaxes
(which are symmetric) and the strongly skewed distribution of true parallaxes.
The net effect is a trend to include, and with more weight, stars with
parallaxes measured too high, rather than stars with parallaxes measured too
low. If not corrected, this bias leads to underestimate the average distance
of the sample. Furthermore, due to classical Malmquist bias, the sample will
likely contain more stars whose magnitudes are erroneously measured too bright
vs stars with magnitude measured too faint. The corresponding average
magnitude will thus be systematically overestimated. Given the small range in
M$_V$ of the subdwarf sample, this effect is small in G97, C98, and the present
paper, while should be accounted for in R97, 98 and PMTV who used also
turn-off stars in their fittings.

The Lutz-Kelker bias affects both the {\it a priori} and the {\it a posteriori}
samples.
\par\noindent
{\it 2. Interstellar reddening}\\ 
The interstellar reddening absorbs light coming from a star, thus weakening
its magnitude, and reddens its color, thus simulating a higher metallicity for
the stars in the {\it a posteriori} sample\footnote{The effect is much smaller
and of opposite sign for stars in the {\it a priori sample}, since
temperatures used in abundance analysis would be underestimated because of
reddening, and, in turn abundances would be underestimated}. Although we have
restricted the samples to stars with very accurate parallaxes (i.e. generally
nearby objects), significant reddening corrections may still be required in a 
few
cases.
\par\noindent
{\it 3. Contamination by binaries}\\
A binary MS companion may brighten and redden the intrinsic $(B-V)_{Mv=6}$
color of the star under consideration, the amount of reddening actually
depending on the magnitude difference between the two components. If the
unresolved binary system is formed by stars with similar masses (and then
luminosity), the primary component brightness can be spuriously enhanced of up
to 0.75 mag (corresponding to the case of strictly equal masses). In turn,
such star will more likely be included in the sample due to Malmquist bias.
\par\noindent
{\it 4. Metallicity bias}\\
As described in Section 4.2 this bias arises from the metallicity distribution
of the stars in the Hipparcos catalogue. This bias requires large corrections;
however it only affects the {\it a posteriori} selected sample\footnote{It
should be mentioned that a small metallicity bias affects also the a priori
sample. In fact, since only stars with measured [Fe/H]$<-0.9$\ are selected,
close to this limit only stars with negative errors are included, and average
metallicities are thus underestimated. However, the net effect is small
because errors in metallicity are small ($\leq 0.15$~dex). The validity of the
adopted interpolating relation between metallicity and MS color was checked 
using stars much more metal-rich than this limit}.

In order to properly take into account and correct for all the above biases,
one should in principle have an exact knowledge of (i) spatial and metallicity
distributions of the calibrating stars, (ii) distribution of measuring errors,
(iii) distribution in mass of the secondary components in binary systems etc.
Since all these quantities are generally poorly known, the safest approach is
to restrict the analysis to (a) only stars with very accurate parallaxes; (b)
very low reddenings; (c) discard binaries; and (d) use the {\it a priori}
sample approach since this sample is not affected by the {\it metallicity
bias}.

The {\it a posteriori} sample may however be used to check the completeness
achieved by the {\it a priori} sample. There are 54 candidate metal-poor (i.e.
$(B-V)_{M_V=6}<0.65$, corresponding to [Fe/H]$<-1.1$ using eq. 4 in Paper
I), unevolved ($5<M_V<8$) stars with good parallaxes (i.e. $\Delta
\pi/\pi<0.12$), in the {\it a posteriori} sample. 22 of these stars have large
errors associated with their $(B-V)$ colors ($\Delta (B-V)>0.05$ mag) and
should be discarded. Most of these objects are known visual binaries not
separated by the Hipparcos beam; colors for these stars cannot be trusted. If
we eliminate all stars with such large color errors, we are left with 32
stars. Reliable metal abundances exist in the literature for 26 of them (but
only 11 are from high dispersion spectroscopy). According to these
determinations 6 of these stars have metallicity [Fe/H]$>-0.9$, thus falling
outside the range defining the metal-poor stars. It is also interesting to
note that while all the 9 objects having [Fe/H]$<-1.5$ are included in the
{\it a priori} sample; $viceversa$, 9 stars of the {\it a priori} sample with
metallicity [Fe/H]$<-1.1$ (from high resolution spectroscopy) are $not$
included in the {\it a posteriori} sample, since their too red measured colors
lead to derive metal abundances [Fe/H]$>-1.1.$

In summary, the {\it a posteriori} sample contains only 6 new good candidate
metal-poor stars (i.e. with [Fe/H]$<-1.1$); 3 of them have $(B-V)_{M_V=6}<
0.63$ (corresponding to [Fe/H]$\leq -1.25$) and only one star has a color
appropriate for [Fe/H]$<-1.5$. Noteworthy, even in this sample there is no
star with [Fe/H]$<-2$.
 
According to these numbers, in the {\it a priori} sample there are 20 stars
with $(B-V)_{M_V=6}< 0.65$ and [Fe/H]$<-1.1$ over a total number of good
candidates found in the {\it a posteriori} sample ranging from 20 to 26
(adding the 6 new good candidates). We can then estimate that the {\it a
priori} sample is from 77 to 100\% complete, the exact percentage depending on
the fraction of the metal-poor candidates that actually are metal-poor stars.
Taking into account the strong asymmetry in the metallicity distribution of
the stars in the solar neighborhood, we estimate that the {\it a priori}
sample is $\sim 90$\% complete.

We conclude that the {\it a priori} approach yields a quite complete sample.
Furthermore, the cleaner definition of the intervening biases and corresponding
corrections, makes it more reliable than the {\it a posteriori} based analysis,
where the noticeable contamination of the sample forces the application of
large and uncertain corrections. In the following Section, we describe the use
of the extended list of unevolved stars with accurate parallaxes and reliable
abundance determinations in the {\it a priori} sample, as fiducial calibrators
to derive new distances for the 9 clusters studied in Paper I. The comparison
with specific points of other studies will be also discussed, where needed.

\subsection{Analysis of the {\it a priori} sample}

Basic data for the new calibrating subdwarfs are described in Section 4.1 and
shown in Table~\ref{t:tab2}. Since stars used in the analysis come from a
larger original population (all stars with [Fe/H]$<-1$, $V<12$, and with
parallax measured by Hipparcos with accuracy $\Delta \pi/\pi<0.12$), and
since a weight proportional to $(\pi/\Delta\pi)^2$ is associated to individual
values, it is necessary to apply Lutz-Kelker corrections. These must be
derived from the properties of the original sample.

\subsubsection{Lutz-Kelker correction}

Following the procedure by Hanson (1979), we used the proper motions in the
Hipparcos catalogue to derive Lutz-Kelker corrections of the parallaxes (i.e.
absolute magnitudes) of our calibrators. The proper motion distribution of the
stars in the {\it a priori} sample is shown in Figure~2. This
follows very well a power law described by the relation $\mu^{-n}$, where $n$
is related to the corresponding parallax distribution ($\sim \pi^{-n}$). For
values of $\mu>0.25$~arcsec/yr (see Figure~2), $n=3.62\pm 0.13$.
Since about the 80$\%$ of the original sample (i.e. the 246 stars from which
the {\it a priori} sample was drawn) is included in this limit of $\mu$ and
since stars with $\Delta \pi/\pi<0.12$ form only $\sim 20$\% of this original
sample, it seems appropriate to assume that parallaxes around this threshold
are distributed as $\pi^{-3.62\pm 0.13}$. This exponent of the parallax
distribution is very close to the value ($n=4$) expected for a uniform
distribution. This implies that the adopted magnitude cut ($V=12$) is not too
much severe (for a comparison, R97 finds $n=3.4$). According to this value of
{\it n}, the absolute magnitudes of our calibrators were corrected using the
relation:
\begin{equation}
\Delta M_{LK}=-8.94\,(\Delta \pi/\pi)^2-63.92\,(\Delta \pi/\pi)^4
\end{equation}
We note anyway that given the severe limit in the parallax error, the
Lutz-Kelker corrections for our sample are however rather small, the maximum
value being $\Delta M_{LK}=0.14$~mag (the average being 0.02~mag).

\subsubsection{Binary contamination}

Among the 56 stars in the {\it a priori} sample having good parallax ($\Delta
\pi/\pi<0.12$), 19 (i.e. $34$\%) are known or suspected binaries. This binary
fraction is smaller than in Paper I ($47$\%: 16 out of 34 stars), but we
believe this is simply due to the lack of accurate observations for most of the
additional stars in this more extended sample. Therefore, a larger fraction of
unknown binaries is expected to contaminate the candidate {\it bona fide}
single stars of the present sample.

We have compared the residual distribution with respect to the
$(B-V)_{Mv=6}-$[Fe/H] relationship (eq. 4 of Paper I) of both known binaries
and {\it bona fide} single stars (in the range $5<M_V<8$) in the extended
sample, in order to estimate the systematic correction required to properly
take into account the binary contamination. According to the procedure devised
in Paper I (see Section 5.1 of that paper), we write this correction as the
product of the probability of a star to be a binary and of the average
correction for each binary. Using the present sample, we derived a probability
of $0.15^{+0.5}_{-0.15}$ for {\it bona fide} single stars to actually be 
binaries (as expected, this figure is larger than found in Paper I), and an
average correction (derived from the average offset of the known binaries with
respect to the $(B-V)_{Mv=6}-$[Fe/H] relation) of $0.13^{+0.10}_{-0.13}$~mag.
Therefore, the appropriate binary correction to apply to {\it bona fide}
single stars in the present {\it a priori} sample is
$0.02^{+0.06}_{-0.02}$~mag.

Although a larger number of binaries are expected in the field than among
cluster stars, where wide, primordial binaries are likely to be destroyed by
interactions with other cluster stars, some contamination by undetected
binaries may still affect the MS loci of the clusters in our sample. Moreover,
blending of stars in the crowded field of GCs may mimic the physical binarity.

A search for binaries in the outer regions of some GCs has led to a binary
occurrence of $\leq 20$\%  (Pryor et al. 1989; Kaluzny et al. 1998; Rubenstein
\& Bailyn 1997; Ferraro et al. 1997), but larger values are found by (Fischer
et al. 1994; Cote et al. 1994) based on radial velocity surveys with
Fabry-Perot spectrographs.

The use of modal instead of mean values to identify the MS mean loci of GCs
(e.g. Sandquist et al. 1996) may help reducing this effect. However, we used a
Monte Carlo simulation and estimate that a residual "reddening" of about $\sim
0.005$~mag (with an uncertainty of about 50\%) might be present in the c-m
diagrams used in our analysis. In turn, distance moduli could be corrected by
$\sim 0.03\pm 0.02$~mag, in the sense to increase distances. However, given
its large uncertainty we did not  apply this correction.

\subsubsection{The $(B-V)_{Mv=6}-${\rm [Fe/H]} relation}

The goodness of the $(B-V)_{Mv=6}-$[Fe/H] derived in Paper I (Equation 4) was
checked using the enlarged sample of subdwarfs of the present {\it a priori}
sample. Only stars with $5<M_V<8$ mag were used to perform this test (30 out
of 56 objects in the sample). Results are illustrated in Figure~3, where
filled symbols indicate {\it bona fide} single stars and open symbols are used
for  known or suspected binaries. Plotted over the data are the theoretical
relations of Straniero \& Chieffi (1991)\footnote{We used [m/H] rather than
[Fe/H], in order to include the 0.3 dex enhancement due to the
$\alpha-$elements, since this was not taken explicitly into account in these
models}.

On average, the {\it bona fide} single stars are $0.005\pm 0.007$ mag
($\sigma=0.033$ mag) redder than the adopted relation  with no clear trend with
metallicity. This small (not very significant) difference could be due to
undetected binaries which contaminate the sample. 

We explicitly note that the observed $(B-V)_{Mv=6}-$[Fe/H] relation must be
extrapolated, when considering the most metal-poor clusters (e.g. M92), since
there are no unevolved stars with good parallax and [Fe/H]$<-2$, in the {\it a
priori} sample. This is not an intrinsic limit of our sample though, since the
comparison with the {\it a posteriori} sample assures that the lack of 
subdwarfs with very good parallaxes and metallicities lower than 
$\sim -$2 dex 
is indeed an intrinsic limitation of the
Hipparcos catalogue\footnote{The apparent presence of more metal-poor stars in
PMTV sample is simply due to their adoption of a lower metal abundance scale.
All of the most metal-poor stars in PMTV sample are also present in our {\it a
priori} sample, but we attribute them a higher metallicity}.

\subsection{Globular Cluster distances and ages using MS fitting results}

The new dataset of field subdwarfs selected {\it a priori} was used to
re-derive distances and ages for the 9 GCs already studied in Paper I. The
same data for the clusters (photometry, reddening, etc.), and the same general
criteria and procedures defined in Paper I were adopted in the present
analysis, with two main differences: i) we used all the stars with $5<M_V<8$
mag; CD$-80^0328$ and HD121004 were discarded though, since they yield very
discrepant results (in opposite directions; we suspect colors or metal
abundances for these two stars to be wrong), and ii) we considered only stars
whose metal abundances were within 0.5 dex from that of the analyzed cluster.
Colors for the subdwarfs were corrected following eq. 4 of Paper I, to account
for differences in metallicity between clusters and calibrating subdwarfs.

Figure~4 shows the best fittings of the cluster mean loci to the fiducial
sequences defined by the local calibrators. Only {\it bona fide} single stars
with $5<M_V<8$ were used in the fits. Relevant results obtained from the
fittings are summarized in Table~\ref{t:tab4}.
 
The absolute ages shown in this table were derived from Straniero, Chieffi \&
Limongi (1997) calibration of the turn-off luminosities. This set of model
isochrones uses a value of $M_{V_\odot}=4.82$ mag, as recommended by Hayes
(1985), and provides ages which fall in the middle of those obtained using
other isochrone sets. Indeed, almost any of the most recent models result into
very similar ages, once, as discussed in Paper I, the same
luminosity-to-magnitude transformation is used. The new determinations compare
very well with the results of Paper I. On average, the new distance moduli are
0.04 mag shorter, and the corresponding ages 0.5 Gyr larger, than in Paper I.
In turn, there is again overall agreement between the present paper, R98 and
C98. In this respect, thanks to R98 adoption of GCC metallicity scale,
the difference in the distance moduli is decreased from $<\Delta (m-M)_V> =
-0.07\pm 0.03$ mag, $r.m.s.=0.08$ mag (6 clusters) of  [Paper I $-$ R97], to
$<\Delta (m-M)_V> = -0.03\pm 0.02$ mag, $r.m.s.=0.05$ mag (5 clusters) of [the
present paper $-$ R98].

\subsection{The case of M92}

\noindent
The GC M92 deserves a deeper discussion since it is often considered 
the prototype of an old GC, and because its distance modulus, as derived by
independent MS fitting analyses, spans a range of about 0.30 mag
(Table~\ref{t:tab1}). The distance modulus we derive here for M92, without
binary correction [$(m-M)_V=14.74$ mag], is identical to the value obtained by
PMTV, again without binary correction. However this
coincidence is totally fortuitous since (i) a different (lower) metallicity,
(ii) a different (higher) reddening, and (iii) a much larger binary correction
is assumed by PMTV for the subdwarf calibrators used to fit the M92 MS.
Indeed, a number of arguments makes the application of the subdwarf fitting
 technique to
M92 difficult and, the corresponding estimated distance and age uncertain:

a) the zero-point of M92 photometry is still uncertain. There is a 0.03
mag difference in the color of the M92  MS, between the photometries of
Heasley \& Christian (1986) and Stetson \& Harris (1988), the latter being
bluer (thus yielding a longer modulus and a younger age). While Stetson \&
Harris photometry seems very reliable, a small zero-point error of $\pm 0.01$
mag may still be present in the colors (Stetson \& Harris 1988). A similar
error may be present in the reddening estimate. The effect of errors in metal
abundances may also be as large as $\pm 0.06$ dex, implying an uncertainty of
$\pm 0.03$ dex in the distance. Combining these facts, the internal
uncertainty on the age for this single cluster amounts to about 2-2.5 Gyr,
that is equal to the difference between the age we derive for M92 (14.8 Gyr)
and the average value for all ``old" clusters (12.3 Gyr).

b) according to GCC and CG there are no subdwarfs with metallicity comparable
to M92 (i.e. [Fe/H]$<-2$) and with reliable parallax in the Hipparcos
catalogue. As mentioned above, this is an intrinsic limitation of the catalogue
itself, and not of our sample. Indeed, R98 did not re-examine M92 and the
other most metal-poor clusters, since Hipparcos measured only two very
metal-poor subdwarf candidates. The claimed presence of stars with good
Hipparcos parallaxes, unevolved and with [Fe/H]$<-2$, in PMTV and R97, is
simply due to their adoption of Carney et al. (1994) metal abundances for the
field subdwarfs. However, as found by CGCS, Carney et al. abundances are too
low (by about 0.36 dex) for the most metal-poor stars. PMTV distance to
M92 is similar to that derived in the present paper only because they use
larger reddenings for the field subdwarfs (from Arenou, Grenon \& Gomez 1992),
and this fortuitously offsets the difference due to metallicity.

In view of all the uncertainties still involved in the M92 analysis, 
the distance to this cluster has a large error bar, and  care should be
exerted before drawing any conclusion on the age of the {\it oldest} GCs from
the analysis of just M92.

\subsection{The impact of the adopted metallicity scale on the Pop. II
distance scale: some caveats}

Due to the strong sensitivity of the subdwarf fitting method to metal
abundance ($0.4<d(m-M)_V/d[{\rm Fe/H}]<1$, the exact figure depending on
[Fe/H], see Paper I for a detailed discussion of this point) the use of a
strictly homogenous metallicity scale for cluster and field stars may still be
not enough to assure derivation of correct cluster distances and ages. Indeed,
although we have used GCC/CG metallicity scale for all stars involved in the 
fittings,
systematic errors may still be present because:

\par\noindent
1) cluster abundances are usually derived from giants while field calibrators
are dwarfs. Our abundances are derived using model atmospheres extracted from
Kurucz (1993) grid. However, these models may better reproduce real
atmospheres of dwarfs than of giants (see Gratton, 1998a, and CGCS99). Thus,
residual systematic differences could possibly exist between abundances
derived for these two types of stars. We estimate that these differences may
be as large as 0.1 dex.

\par\noindent
2) As well known, metal-poor stars are expected to show a substantial
enhancement of the abundances of $\alpha-$elements with respect to iron,
([O/Fe]$\sim 0.45$, [$\alpha$/Fe]$\sim 0.3$) due to the interplay of the
evolutionary times of SN~I and SN~II progenitors. However, two recent papers
revealed that there may be exceptions. Carney et al. (1997) derived
[Mg/Fe]=$-0.31$ for BD$+80^0 245$ ([Fe/H]=$-1.86$), and King (1997) found
[Mg/Fe]=$-0.10$ for the common proper motion pair HD134439/134440
([Fe/H]=$-1.50$). Since distance determinations depend on the [$\alpha$/Fe]
ratio (being $0.3<d(m-M_V)/d[\alpha/{\rm Fe}]<0.7$, from Straniero \& Chieffi
1991), the problem is to establish how large is the fraction of
$\alpha-$underabundant, metal-poor stars.

The underabundance in field stars could possibly be correlated with
kinematics, as suggested by the large apogalactic distance of HD134439/134440.
We have used an independent sample of 39 stars with [Fe/H]$<-1$ and [Mg/Fe]
from Fuhrmann, Axer \& Gehren (1995), and Zhao \& Magain (1990), to test the 
actual existence of this kinematics induced underabundance. We found that only 
a small minority of the metal-poor stars have $\alpha-$underabundance, with no
clear correlation with kinematics. The lack of $\alpha-$enhancement in a few
halo stars should then have only very minor effects on distance derivations.

As a further check of the impact of anomalous $\alpha-$abundances in local
subdwarfs, we have derived the distance modulus of the programme clusters
eliminating the couple HD134439/HD134440 from the sample (BD$+80^0 245$\ is
not amongst our calibration subdwarfs). Changes are very small: on average,
distance moduli are decreased by $0.004\pm 0.005$~mag, with no corrections
larger than 0.02~mag. Such changes will be neglected hereinafter.

\subsection{Residual overall uncertainties in the subdwarf fitting 
distances and ages for Globular Clusters: future directions}

The applicability of the subdwarf fitting technique rests on the assumption
that the metal-poor field subdwarfs in the solar neighborhood are the local
counterpart of the GC MS stars. Since a direct measure of the trigonometrical
parallax of GC stars is not in the reach of present day instrumental
capabilities [but will became feasible with the accomplishment of the GAIA
(Lindgren \& Perryman, 1996) mission] we must rely on this assumption.
However, for a robust application of the method, field calibrators and cluster
stars should be analyzed in a consistent way. Indeed, in our procedure we made
a serious effort to reduce any systematics arising from possible differences
in the treatment of  field subdwarfs and GC MS stars. However, some residual
uncertainties still exist. In table~\ref{t:tab5} we summarize the most
relevant sources of uncertainties present in our application of the subdwarf
fitting technique.

Since we only used subdwarfs (i) with a limited range in magnitude ($5<M_V<8$),
(ii) with very accurate parallaxes ($\Delta \pi/\pi < 0.12$), and (iii) {\it a
priori} selected, the statistical biases (Malmquist bias, Lutz-Kelker,
metallicity bias) contribute only marginally ($\sim$ 0.02 mag) to the final
error on the derived distance moduli. Contamination by binaries among field
subdwarfs (when only {\it bona fide} single stars are used), or in the cluster
MSs (when modal instead of mean is used to define the loci sequence loci),
contribute an additional $\sim$ 0.04 mag.

Presently, the largest sources of uncertainty reside with (i) colors and
photometric calibrations, (ii) reddening, and (iii) metal abundances and
metallicity scale, for both the subdwarf calibrators and the GC MS stars.

Due to the steepness of the MS, small errors in the photometry may cause large
errors in the derived distance moduli. Furthermore, at present, the only color
suited for the subdwarf fitting is the $B-V$, since reliable $V-R$\ and $V-I$\
colors are lacking for many of the field subdwarfs (see CGCS99) and
near-infrared deep color magnitude diagrams are not available for most
clusters.

To avoid systematic effects, field calibrators and cluster stars should be
observed with the same instrumental configuration and in the same photometric
system. However, given the large difference in luminosity between the two
samples (about 10 mag) this is not feasible. Photomultipliers and filters for
the Johnson-Cousins system are generally used for the field calibrators, while
CCD's are used for the GCs. Magnitudes for the latter are then transformed to
Johnson-Cousins by observation of Landolt's standards (Landolt 1983, 1992).
Albeit considerable care is generally devoted to the calibrations of the GCs
data, some uncertainty still exists, and results for individual clusters may
well have rather large errors. We estimate a total photometric uncertainty of
$\sim$ 0.04 mag (distance modulus) as the average over the 9 clusters
considered here.

A consistent reddening scale should be used for cluster and dwarf stars. Up to
now, a direct comparison of the excitation temperatures of field subdwarfs and
GC MS stars has not been feasible (important progresses are expected from UVES
at VLT2); however some constraints can be derived by comparing the reddenings
adopted for GCs and template subdwarfs with a cosecant law. Reddenings for the
GCs were taken from Zinn (1980), this procedure is common to R97,98, G97 and
PMTV. Two different reddening scales are available, instead for the subdwarfs.
In the present analysis we adopt reddenings from Carney et al. (1994),
Schuster \& Nissen (1989), and Ryan \& Norris (1991). A star-by-star
comparison shows that they are on a uniform scale. PMTV reddening for the
subdwarfs were instead taken from Arenou et al. 1992. When compared to
cosecant-laws for reddening (Bond 1980), GCs and subdwarfs are on a uniform
reddening scale if the height scale of the galactic dust disk is 100~pc in our
reddening scale for the subdwarfs, and 40 pc if PMTV scale is adopted. While
the former value is in the middle of current determinations of the galactic
dust scale-height (50-150 pc: Lynga 1982, Pandey \& Mahra 1987, Scheffler \&
Els\"asser 1987, Spitzer 1978, Salomon et al. 1979, Burton 1992, Chen 1998),
the latter is at the lower extreme of the admitted range.

On the whole, we think that the reddening scale of subdwarfs still carries an
uncertainty of $\pm 0.015$~mag; this translates into an uncertainty of $\sim
0.07$ mag in the derived distance moduli.

Finally, differences in the adopted metallicities of subdwarfs and clusters
stars may still be present even if a homogeneous metallicity scale was
adopted, since abundance for clusters are derived from giants instead of MS
stars. For this reason systematic differences (of $\sim$0.1 dex) may exist
between the metallicity of the template subdwarfs and the GC stars. This
translates into a corresponding uncertainty of $\pm$0.08 mag in the derived
distance moduli.

As an example, Figure~5 displays the best fit of the 9 clusters analyzed in
the present paper obtained assuming that CG metallicity for these clusters is
underestimated by 0.1 dex. The fitting of the cluster subgiant branches seem
to be slightly improved. However, a similar effect could also be produced by
errors in the adopted reddenings for the few subgiants with good parallaxes
used in the fits, or if the cluster reddening scale were systematically
underestimated by $\sim 0.01$~mag, well within the uncertainties of current
reddening estimates for the clusters.

Adding up in quadrature the errors listed in Table~\ref{t:tab5} we obtain 0.12
mag as total error budget associated with our subdwarf fitting distances to
GCs. The analogous figure quoted in Table 2 of Renzini (1991) was 0.25 mag. At
the time the Renzini's paper was written, this large error was mainly accounted
for by uncertainties in the parallax data. Thanks to Hipparcos this
uncertainty has now been more than halved, and no longer resides with
parallaxes, but with photometric calibrations, reddening and metallicity
scales.

\section{THE $M_V$(HB)$-$[Fe/H] RELATION AND DISTANCE MODULUS OF THE LMC}

\subsection{Results from subdwarf fitting method}

Using data listed in Table~\ref{t:tab4}, we derive the following weighted best
fitting relations between absolute magnitude of the HB and metallicity:
\begin{equation}
M_V(HB)   = (0.13\pm 0.09) ({\rm [Fe/H]}+1.5) + (0.44\pm 0.04)~[\pm 0.12]
\end{equation}
\begin{equation}
M_V(ZAHB) = (0.18\pm 0.09) ({\rm [Fe/H]}+1.5) + (0.53\pm 0.04)~[\pm 0.12]
\end{equation}
\begin{equation}
M_V(RR)   = (0.18\pm 0.09) ({\rm [Fe/H]}+1.5) + (0.47\pm 0.04)~[\pm 0.12]
\end{equation}
Here 0.04 mag is the internal error, however as discussed in Section 4.8 a
more realistic estimate of the error is 0.12 mag. 

We recall that $M_V(HB)$ is the average magnitude of the HB (see also
Section 5.2.3 of Paper I). Due to evolution, this luminosity is somewhat
brighter than the Zero Age HB luminosity $M_V(ZAHB)$, usually provided by
theoretical models. Following Paper I we used Sandage (1993) relation to tie
$M_V(HB)$ to $M_V(ZAHB)$. $M_V(HB)$ does not coincide either with the average
magnitude of the RR Lyrae variables, $M_V(RR)$. We adopted Caloi et al. (1997)
correction between $M_V(ZAHB)$ and $M_V(RR)$.

In Paper I we estimated the age of the oldest GCs by considering only
Oosterhoff II and Blue HB Clusters. If, however, we are rather interested in
the epoch of formation of GCs, all the nine clusters of Table~\ref{t:tab4}
should be considered (we incidentally note  that deviations of individual
clusters from the average value are of the same order of the expected accuracy
of the internal errors of the MSF technique: more accurate relative ages can
be obtained using other techniques). A simple average of the ages in
Table~\ref{t:tab4} gives $12.2\pm 0.5$~Gyr ($\sigma=1.4$~Gyr r.m.s. of values
for individual clusters). However, as discussed in Paper I, this error bar (as
well as the simple mean value) is incorrect, because (i) systematic errors are
much larger than random errors; and (ii) some of the error bars are not
symmetric (for instance, uncertainties in the consideration of diffusion may
only lead to reducing the ages of Table~\ref{t:tab4} which were derived
neglecting diffusion). Following the approach of Paper I, we then estimated a
more realistic error bar using a MonteCarlo technique. Table~\ref{t:tab6}
lists the individual sources of errors considered in this simulation, as well
as the type of distribution and the adopted parameters\footnote{Among the
other source of errors, we mention here that we assumed a flat distribution
for [$\alpha$/Fe] between +0.2 and +0.5~dex, with a preferred value of
+0.3~dex. We believe this is the best estimate from current determinations
in the metallicity range which is relevant in the present context :
$-2<$[Fe/H]$<-1$. Adopting Chaboyer
et al. (1998)  age - [$\alpha$/Fe] dependence 
the corresponding error in ages ranges from $-$0.8 to
+0.3~Gyr}. The resulting mean age for the nine GCs is:
\begin{equation}
{\rm Age}=11.5\pm 2.6~~~~{\rm Gyr}
\end{equation}
(95\% confidence range). While this value nearly coincides with that of Paper
I, its meaning is different. In fact, in Paper I that was the age of the
oldest GCs, here it represents the mean age of the galactic GCs. If the same
cluster selection made in Paper I is made, we would derive an age of $11.9\pm
2.7$~Gyr.

Finally, using eq. (4) we derive the following estimates for the LMC distance
modulus: 
$$\mu_{\rm LMC}=18.64\pm 0.04~[\pm 0.12]$$

for an adopted average dereddened magnitude of the RR Lyraes in the bar of 
the LMC $<V_0>=19.11\pm 0.02$.

\subsection{Best estimate of the distance to the LMC and of the 
age of globular clusters}

The LMC is widely considered a corner-stone of the astronomical distance
scale. In Table~\ref{t:tab7} we have summarized different estimates of the
distance modulus of the LMC obtained by a large number of independent
techniques, some of which based on Hipparcos parallaxes (see Section 3). 
Our subdwarf fitting modulus for the LMC is slightly larger than
those provided by the other techniques: however, {\bf within their error bars
almost all methods provide the same distance to the LMC. This result is new,
and stems from the adoption of a consistent reddening scale in the various
distance determinations. Once the same reddening scales are adopted for
Cepheids and RR Lyrae, the short and long distance scale agree within their
nominal error bar.} The only result which is clearly discrepant at present
is that from the eclipsing binary HV2274 by Guinan et al.; if we accept its
internal error as a realistic estimate of the true error bar, this result is
more than 3~$\sigma$ from the average of all other determinations. This might
indicate that some problem still exists on the distance to the LMC; however,
we (aribratirily) prefer to wait more data on eclipsing binaries in the LMC
before assigning weight to this method.

If the distance to HV2274 is neglected, the (weighted average) distance
modulus of the LMC is:
$$\mu_{\rm LMC}=18.54\pm 0.03\pm 0.06,$$
where the first error bar is simply the 1~$\sigma$\ error bar due to scatter
in the individual determinations, while the second one derives from an 
(arbitrarily) assumed uncertainty of $\pm 0.02$~mag in the reddening scale.

The main implications of this result are:
\begin{enumerate}
\item The cosmic distance scale is $2\pm 2$\% longer than previously assumed,
and the value of the Hubble constant correspondingly smaller. The change 
with respect to the value usually assumed in extragalactic studies is small
and may be reasonably neglected for most purposes.
\item The $M_V(HB)$--[Fe/H] relations implied by this distance scale are:
\begin{equation}
M_V(HB)   = (0.13\pm 0.09) ({\rm [Fe/H]}+1.5) + (0.54\pm 0.07)
\end{equation}
\begin{equation}
M_V(ZAHB) = (0.18\pm 0.09) ({\rm [Fe/H]}+1.5) + (0.63\pm 0.07)
\end{equation}
\begin{equation}
M_V(RR)   = (0.18\pm 0.09) ({\rm [Fe/H]}+1.5) + (0.57\pm 0.07)
\end{equation}
\item The ages of the GCs derived assuming this distance scale (which is
1~$\sigma$\ shorter than given by the subdwarf fitting method) are 1.4 Gyr
older than found in Section 5, well within the quoted error bar. The average
age of the GC would then be:
\begin{equation}
{\rm Age} = 12.9\pm 2.9~~~{\rm Gyr}
\end{equation}
(95\% confidence range).
\end{enumerate}

\section{THE EPOCH OF FORMATION OF GLOBULAR CLUSTERS AND THE COSMIC SFR}

The Milky Way provides basic data to study the cosmic evolution of galaxies.
However, up to a few years ago, the discrepancy existing between the age
estimated for the galactic GCs and for the present and past expansion rate of
the Universe hampered the use of this template. In particular, it was
impossible to compare the epoch of formation of our own Galaxy with evidences
from high redshift objects.

While large uncertainties still exist, the situation has now changed thanks to
(i) the very recent results about the Universe de-acceleration parameters
obtained by type Ia SNe studies (Schmidt et al. 1998; Pearlmutter et al. 1998;
Garnavich et al. 1998), and to (ii) the revised estimates of the age of the
(old) galactic GCs now possible thanks to Hipparcos. In Figure~6, we have
plotted the redshift of the formation of GCs against various possible values
for the Hubble Constant. The three panels show the results obtained for three
different values of  $\Omega_M$\ (0.2, 0.3, and 0.4), within a flat Universe
($\Omega_\Lambda=1-\Omega_M$). The solid line shows the favourite age of GCs
according to the present paper (12.9~Gyr), and the dashed lines represent the
limits of our 95\% level of confidence. While the admitted area is still
large, it is possible to locate at quite high confidence the epoch of
formation of galactic GCs (that we identify with the first major episode of
star formation in our Galaxy, corresponding to the bulk of the halo, the thick
disk, and perhaps the bulge) at $z>1$\  the preferred value being at $z\sim 3$.

This value for the epoch of formation of GCs compares well with evidences from
high redshift galaxies, as given by analysis of the HDF, which locates the
bulk of cosmic star formation at $z\simeq 1$\ (Madau et al. 1998). An older
star formation (at $1<z<4$) was obtained by Franceschini et al. (1998) for
elliptical galaxies in the HDF (assuming $q_0=0.15$, roughly corresponding to
$\Omega_M=0.3$). This comparison is perhaps more relevant here, since our age
estimate does not include the thin disk of the Milky Way, whose formation
likely occurred later (see e.g. discussion in Gratton et al. 1996).

\section{CONCLUSIONS}

The literature of the last year has seen the flourishing of a number of new
techniques to measure distances as well as a re-newed interest in the
classical methods which have been revised in light of Hipparcos data (see
Section 3).

A lively debate is taking place among authors who favour one method to the
other and, in turn, one distance scale to the other. By significantly
increasing the number of local subdwarfs with accurate parallaxes, Hipparcos
has allowed to definitely improve the subdwarf MS fitting technique. Work
still remains to be done, though. In fact, while waiting for the benefits of
the next generation astrometric missions (GAIA, SIM see
http://sim.jpl.nasa.gov/sim/), all the efforts should be devoted to cutting
down the 0.12 mag residual uncertainty still affecting the MSF distances to
GCs. New, deep, and precisely absolute-calibrated photometric data should be
collected for GCs (M92 in particular), reddening determinations should be
improved, and abundance analysis of cluster MS stars should be performed.

Based on our detailed analysis, the MS fitting method favours 
the {\it long} distance scale and provides a distance modulus 
for the LMC of : $\mu_{\rm LMC}=18.64\pm 0.04~[\pm 0.12]$, (based on
 $<V_0>=19.11\pm 0.02$\ for the  RR Lyraes in the bar of the LMC), 
and an average
age for the 9 analyzed clusters of ${\rm Age}=11.5\pm 2.6~~~~{\rm Gyr}$.

Our subdwarf fitting modulus for the LMC is 1 $\sigma$ larger than 
provided by other techniques. However, whithin their error bars all methods
lead to the same distance to the LMC once the same reddening scale is 
adopted for Cepheid and RR Lyrae variables.  
Short and long distance scales may be reconciled on an
average distance modulus for the LMC bar of :

$$\mu_{\rm LMC}=18.54\pm 0.03 \pm 0.06$$

The corresponding average age of the GCs would than be :

$${\rm Age}=12.9\pm 2.9~~~~{\rm Gyr}$$

\acknowledgments
{We wish to thank Dr. P.L. Bernacca and M. Lattanzi for their help to access
the Hipparcos data, and Dr. C. Corsi and C. Sneden for their collaboration at
an early stage of this research. We also thank M. Bolzonella, V. Castellani,
S. Degl'Innocenti, M. Feast, M. Marconi, L. Moscardini and M. Zoccali for
helpful and pleasant discussions. The financial support of Agenzia Spaziale
Italiana (ASI) and of Ministero della Universita' e della Ricerca Scientifica
e Tecnologica (MURST) is gratefully acknowledged.

This research has made use of the SIMBAD data base, operated at CDS,
Strasbourg, France.}

\scriptsize
\noindent
\begin{table}[ht]
\caption{Hipparcos-based distance moduli for globular clusters}
\begin{tabular}{rllllrlcrl}
\tableline
\tableline
NGC &Other& [Fe/H]& adopted         &      range      &No.  &  $(m-M)_V$      & 
$(m-M)_V$  & cmd   & Study \\
    &     &       & $E(B-V)$        &                 &Stars&  original       & 
bin. corr. & source&       \\
\tableline
6341& M92 &$-2.15a$&$0.025\pm 0.005$&$-2.5 \div -1.5$ & 2&$14.82\pm 0.08$
&14.80&  1   & Paper I   \\
    &     &$-2.15a$&$0.025\pm 0.005$&$-2.5 \div -1.5$ & 4&$14.74\pm 0.07$ 
&14.72&  1   & this paper\\
    &     &$-2.20b$&$0.02$          &$-2.6 \div -1.8$ &17&$14.74\pm 0.05$
&14.67&  1   & PMTV      \\
    &     &$-2.24b$&$0.02$          & $< -1.7$        &10&$14.99\pm 0.10$
&     &  1   & R97       \\
7078& M15 &$-2.15a$&$0.09$          & $< -1.7$        &10&$15.66\pm 0.10$
&     & 12,13& R97       \\
4590& M68 &$-1.95a$&$0.040\pm 0.010$&$-2.5 \div -1.5$ & 2&$15.33\pm 0.08$
&15.31&  2   & Paper I   \\
    &     &$-1.95a$&$0.040\pm 0.010$&$-2.5 \div -1.5$ & 7&$15.27\pm 0.06$
&15.25&  2   & this paper\\
    &     &$-2.09b$&$0.05$          & $< -1.7$        &10&$15.45\pm 0.10$
&     &  2   & R97       \\
7099& M30 &$-1.88a$&$0.039\pm 0.001$&$-2.5 \div -1.3$ & 3&$14.96\pm 0.08$
&14.94& 3,4  & Paper I   \\
    &     &$-1.88a$&$0.039\pm 0.001$&$-2.5 \div -1.3$ & 8&$14.90\pm 0.05$
&14.88& 3,4  & this paper\\
    &     &$-2.13b$&$0.05$          &$< -1.7$         &10&$15.11\pm 0.10$
&     &  4   & R97       \\
6397&     &$-1.82a$&$0.19$          &$-2.05\div -1.5$ & 8&$12.83\pm 0.15$
&     &  9   & R98       \\
6205& M13 &$-1.41a$&$0.020\pm 0.000$&$-1.8 \div -1.0$ & 9&$14.47\pm 0.07$
&14.45&  5   & Paper I   \\
    &     &$-1.41a$&$0.020\pm 0.000$&$-1.8 \div -1.0$ &17&$14.46\pm 0.04$
&14.44&  5   & this paper\\
    &     &$-1.65b$&$0.02$          &$-1.85\div -1.4$ &11&$14.54\pm 0.10$
&     & 5,11 & R97       \\
    &     &$-1.39a$&$0.02$          &$-1.65\div -1.15$& 9&$14.51\pm 0.15$
&     & 5,11 & R98       \\
    &     &$-1.58$ &$0.02$          &$-1.76\div -1.23$& 7&$14.47\pm 0.09$
&     &  5   & C98       \\
6752&     &$-1.43a$&$0.035\pm 0.005$&$-1.8 \div -1.0$ & 9&$13.34\pm 0.07$
&13.32&  6   & Paper I   \\
    &     &$-1.43a$&$0.035\pm 0.005$&$-1.8 \div -1.0$ &18&$13.34\pm 0.04$
&13.32&  6   & this paper\\
    &     &$-1.54b$&$0.02$          &$-1.85\div -1.4$ &11&$13.23\pm 0.10$
&     &  6a  & R97       \\
    &     &$-1.42a$&$0.04$          &$-1.27\div -1.2$ &12&$13.28\pm 0.15$
&     &  6a  & R98       \\
    &     &$-1.51 $&$0.04$          &$-1.76\div -1.23$& 7&$13.33\pm 0.09$
&     &  6   & C98       \\
 362&     &$-1.12a$&$0.056\pm 0.003$&$-1.6 \div -0.8$ & 6&$15.06\pm 0.08$
&15.04&  7   & Paper I   \\
    &     &$-1.12a$&$0.056\pm 0.003$&$-1.6 \div -0.8$ &13&$15.00\pm 0.05$
&14.98&  7   & this paper\\
5904& M5  &$-1.10a$&$0.035\pm 0.005$&$-1.6 \div -0.8$ & 7&$14.62\pm 0.07$
&14.60&  8   & Paper I   \\
    &     &$-1.10a$&$0.035\pm 0.005$&$-1.6 \div -0.8$ &13&$14.59\pm 0.05$
&14.57&  8   & this paper\\
    &     &$-1.40b$&$0.03$          &$-1.6 \div -1.25$& 8&$14.54\pm 0.10$
&     &  8   & R97       \\
    &     &$-1.10a$&$0.02$          &$-1.35\div -0.9$ & 9&$14.58\pm 0.15$
&     &  8   & R98       \\
    &     &$-1.17 $&$0.03$          &$-1.23\div -1.07$& 4&$14.51\pm 0.09$
&     &  8   & C98       \\
 288&     &$-1.05a$&$0.033\pm 0.007$&$-1.6 \div -0.8$ & 6&$14.96\pm 0.08$
&14.94&  9   & Paper I   \\
    &     &$-1.05a$&$0.033\pm 0.007$&$-1.6 \div -0.8$ &12&$14.97\pm 0.05$
&14.95&  9   & this paper\\
    &     &$-1.07a$&$0.01$          &$-1.30\div -0.85$& 9&$15.03\pm 0.15$
&     &  9   & R98       \\
 104&47Tuc&$-0.67a$&$0.055\pm 0.007$&$-1.3 \div -0.5$ & 8&$13.64\pm 0.08$
&13.62& 10   & Paper I   \\
    &     &$-0.67a$&$0.055\pm 0.007$&$-1.3 \div -0.5$ & 7&$13.57\pm 0.09$
&13.55& 10   & this paper\\
    &     &$-0.70a$&$0.04$          &$-0.9 \div -0.45$& 9&$13.68\pm 0.15$
&     & 10   & R98       \\
6838& M71 &$-0.70a$&$0.28$          &$-0.9 \div -0.45$& 9&$14.06\pm 0.15$
&     & 14   & R98       \\
\end{tabular}
\tablerefs{\scriptsize Metallicity sources: a. Carretta \& Gratton (1997) 
b. Zinn \& West (1984), and further updates.\\
CMD sources: 1. Stetson \& Harris (1988) 2. McClure et al. (1987)
3. Bolte (1987b) 4. Richer, Fahlman \& VandenBerg (1988) 5. Richer
\& Fahlman (1986) 6. Penny \& Dickens (1986) corrected according to
VandenBerg, Bolte \& Stetson (1990) 
7. Bolte (1987a) corrected according to 
VandenBerg et al. (1990) 8. Sandquist et al. (1996) 9. Buonanno,
Corsi \& Fusi Pecci (1989) 10. Hesser et al. (1987) 11. Sandage (1970) 
12. Durrel \& Harris (1993)
13. Fahlman, Richer \& VandenBerg (1985)  
14. Hodder et al. (1992)} 
\label{t:tab1}
\end{table}

\noindent
\begin{table}[ht]
\caption{Stars with $\Delta \pi/\pi<0.12$ in the {\it a priori} sample}
\scriptsize
\begin{tabular}{rlrrrrrrccccc}
\tableline
\tableline
   HD   & Gliese  &$E(B-V)$& $V$ & $\pi$ &$\Delta \pi/\pi$& M$_V$ & 
$\sigma_{M_V}$ & $(B-V)_0$ & $\sigma_{(B-V)_0}$ 
&[Fe/H] & $\sigma_{\rm [Fe/H]}$ &\\ 
\tableline
-35 0360& G269-87 & 0.000 &10.253&  16.28& 0.108&  6.20& 0.22& 0.763& 0.019& 
$-$0.90& 0.11&   \\
    6755& G243-63 & 0.030 & 7.632&   7.74& 0.118&  1.94& 0.24& 0.680& 0.019& 
$-$1.48& 0.15&   \\
-61 0282&         &-0.000 &10.109&  11.63& 0.102&  5.34& 0.21& 0.526& 0.019& 
$-$0.96& 0.11&   \\
   10607&         & 0.010 & 8.298&  14.00& 0.053&  4.00& 0.11& 0.562& 0.020& 
$-$1.13& 0.10&   \\
   17072&         &-0.000 & 6.589&   7.57& 0.067&  0.94& 0.14& 0.660& 0.019& 
$-$1.34& 0.15&   \\
   19445& G037-26 & 0.002 & 8.045&  25.85& 0.044&  5.09& 0.09& 0.458& 0.020& 
$-$1.91& 0.07&   \\
+66 0268& G246-38 & 0.000 & 9.916&  17.58& 0.087&  6.07& 0.18& 0.652& 0.020& 
$-$1.92& 0.07& AB\\
   23439& G095-57A& 0.000 & 7.828&  40.83& 0.055&  5.86& 0.12& 0.778& 0.019& 
$-$0.97& 0.11& SB\\
   25704&         & 0.000 & 8.118&  19.02& 0.046&  4.49& 0.10& 0.553& 0.020& 
$-$0.93& 0.07& AB\\
   25329& Gl 158  & 0.000 & 8.502&  54.14& 0.020&  7.17& 0.04& 0.864& 0.018& 
$-$1.69& 0.07&   \\
  284248& G008-16 &-0.008 & 9.257&  12.84& 0.104&  4.70& 0.21& 0.458& 0.020& 
$-$1.57& 0.07&   \\
   29907&         & 0.000 & 9.883&  17.00& 0.058&  6.01& 0.12& 0.632& 0.020& 
$-$1.71& 0.15& S?\\
   31128&         & 0.010 & 9.095&  15.57& 0.077&  5.00& 0.16& 0.480& 0.020& 
$-$1.86& 0.10&   \\
   34328&         & 0.003 & 9.436&  14.55& 0.069&  5.21& 0.15& 0.478& 0.020& 
$-$1.44& 0.07&   \\
   46663&         & 0.010 & 9.514&  21.80& 0.110&  6.09& 0.23& 0.927& 0.020& 
$-$2.11& 0.15& AB\\
-33 3337&         & 0.020 & 9.016&   9.11& 0.111&  3.69& 0.23& 0.452& 0.020& 
$-$1.33& 0.07&   \\
   64090& G090-25 & 0.000 & 8.276&  35.29& 0.029&  6.01& 0.06& 0.614& 0.020& 
$-$1.48& 0.07& SB\\
-80 0328& Gl 345  & 0.012 &10.089&  16.46& 0.060&  6.14& 0.13& 0.553& 0.019& 
$-$1.75& 0.11&   \\
   84937& G043-03 & 0.009 & 8.300&  12.44& 0.085&  3.71& 0.18& 0.382& 0.020& 
$-$2.07& 0.07& SB\\
   89499&         & 0.010 & 8.609&   8.93& 0.082&  3.30& 0.17& 0.687& 0.020& 
$-$1.91& 0.15& SB\\
   91345&         & 0.010 & 9.016&  17.70& 0.053&  5.23& 0.11& 0.550& 0.018& 
$-$0.98& 0.10&   \\
   94028& G058-25 &-0.000 & 8.221&  19.23& 0.059&  4.61& 0.12& 0.482& 0.020& 
$-$1.32& 0.07& SB\\
   97320&         & 0.010 & 8.145&  17.78& 0.043&  4.39& 0.09& 0.447& 0.020& 
$-$1.01& 0.10&   \\
  102200&         & 0.001 & 8.751&  12.45& 0.096&  4.14& 0.20& 0.438& 0.020& 
$-$1.22& 0.15&   \\
+51 1696& G176-53 & 0.000 & 9.913&  13.61& 0.113&  5.46& 0.23& 0.552& 0.020& 
$-$1.26& 0.07&   \\
  103095& G122-51 & 0.000 & 6.422& 109.21& 0.007&  6.61& 0.02& 0.752& 0.019& 
$-$1.24& 0.07&   \\
  108177& G013-35 & 0.002 & 9.662&  10.95& 0.118&  4.72& 0.24& 0.430& 0.020& 
$-$1.55& 0.07&   \\
  111980&         & 0.010 & 8.338&  12.48& 0.111&  3.70& 0.23& 0.532& 0.020& 
$-$1.16& 0.11& AB\\
  113083&         & 0.010 & 8.014&  18.51& 0.061&  4.32& 0.13& 0.540& 0.020& 
$-$1.09& 0.11&   \\
  116064&         & 0.010 & 8.780&  15.54& 0.093&  4.66& 0.19& 0.440& 0.019& 
$-$1.86& 0.07&   \\
  120559&         & 0.020 & 7.918&  40.02& 0.025&  5.92& 0.05& 0.642& 0.019& 
$-$0.95& 0.11&   \\
  121004&         & 0.010 & 8.999&  16.73& 0.081&  5.06& 0.17& 0.606& 0.020& 
$-$0.90& 0.11&   \\
\tableline
\end{tabular}
\label{t:tab2}
\end{table}

\begin{table}[ht]
\normalsize
{\rm Table} 2. \hspace{6.5cm}{\it continued}\\
\scriptsize
\begin{tabular}{rlrrrrrrccccc}
\tableline
\tableline
   HD   & Gliese  &$E(B-V)$& $V$ & $\pi$ &$\Delta \pi/\pi$& M$_V$ & 
$\sigma_{M_V}$ & $(B-V)_0$ & $\sigma_{(B-V)_0}$ 
&[Fe/H] & $\sigma_{\rm [Fe/H]}$ &\\ 
\tableline
  126681&         &-0.001 & 9.302&  19.16& 0.075&  5.66& 0.16& 0.602& 0.020& 
$-$1.09& 0.07&   \\
  132475&         & 0.037 & 8.439&  10.85& 0.105&  3.51& 0.22& 0.522& 0.020& 
$-$1.73& 0.09&   \\
  134440&         & 0.005 & 9.418&  33.68& 0.050&  7.03& 0.11& 0.845& 0.018& 
$-$1.28& 0.07&   \\
  134439&         & 0.005 & 9.052&  34.14& 0.040&  6.70& 0.08& 0.767& 0.019& 
$-$1.30& 0.07&   \\
  140283& GJ1195  & 0.024 & 7.137&  17.44& 0.056&  3.32& 0.12& 0.463& 0.019& 
$-$2.40& 0.07& IR\\
  145417&         & 0.010 & 7.496&  72.75& 0.011&  6.80& 0.02& 0.805& 0.019& 
$-$1.64& 0.11&   \\
 149414A& G017-25 & 0.010 & 9.581&  20.71& 0.072&  6.11& 0.15& 0.726& 0.018& 
$-$1.14& 0.07& SB\\
  159482& G139-48 & 0.020 & 8.320&  20.89& 0.056&  4.89& 0.12& 0.560& 0.020& 
$-$1.06& 0.10&   \\
+05 3640& G140-46 & 0.010 &10.350&  17.00& 0.112&  6.38& 0.23& 0.732& 0.020& 
$-$0.78& 0.09& IR\\
  166913&         & 0.010 & 8.191&  16.09& 0.065&  4.18& 0.14& 0.441& 0.020& 
$-$1.44& 0.07&   \\
  188510& G143-17 & 0.001 & 8.830&  22.80& 0.061&  5.58& 0.13& 0.598& 0.020& 
$-$1.37& 0.07& SB\\
  189558&         & 0.010 & 7.703&  14.76& 0.075&  3.50& 0.16& 0.565& 0.020& 
$-$1.04& 0.10&   \\
+42 3607& G125-64 & 0.040 & 9.986&  12.02& 0.094&  5.30& 0.20& 0.470& 0.020& 
$-$1.79& 0.11&   \\
  193901&         & 0.003 & 8.644&  22.88& 0.054&  5.41& 0.11& 0.549& 0.020& 
$-$1.00& 0.07&   \\
  194598& G024-15 & 0.003 & 8.335&  17.94& 0.069&  4.56& 0.15& 0.484& 0.020& 
$-$1.02& 0.07&   \\
  196892&         & 0.001 & 8.244&  15.78& 0.077&  4.18& 0.16& 0.497& 0.020& 
$-$1.04& 0.15&   \\
+41 3931& G212-07 & 0.030 &10.182&  14.24& 0.103&  5.85& 0.21& 0.584& 0.020& 
$-$1.49& 0.15&   \\
  201891&         & 0.003 & 7.367&  28.26& 0.036&  4.61& 0.08& 0.514& 0.019& 
$-$0.97& 0.07&   \\
  204155& G025-29 & 0.000 & 8.492&  13.00& 0.085&  4.00& 0.18& 0.571& 0.020& 
$-$0.98& 0.10&   \\
-00 4234& G026-09 & 0.010 & 9.765&  20.26& 0.099&  6.21& 0.20& 0.947& 0.019& 
$-$1.09& 0.15& SB\\
  205650&         &-0.000 & 9.044&  18.61& 0.066&  5.35& 0.14& 0.518& 0.020& 
$-$1.00& 0.15&   \\
+59 2407& G231-52 & 0.050 &10.115&  15.20& 0.080&  5.96& 0.17& 0.580& 0.021& 
$-$1.62& 0.15&   \\
  211998&         & 0.010 & 5.280&  34.60& 0.060&  2.91& 0.13& 0.669& 0.020& 
$-$1.43& 0.10& AB\\
 219175A& G157-32 & 0.000 & 7.570&  26.52& 0.091&  4.61& 0.19& 0.544& 0.018& 
$-$1.31& 0.15&   \\
\tableline
\end{tabular}
\end{table}

\small
\noindent
\begin{table}[ht]
\caption{Revised distances and ages for the 9 programme globular clusters}
\begin{tabular}{ccccccc}
\tableline
\tableline
Cluster&Stars&   $(m-M)_V$ &$(m-M)_V$&  $M_V$(HB)    &$M_V$(TO)&Age \\
       &     &   original  &bin cor. &               &         &(Gyr)\\
\tableline
M92   &  4 & $14.74\pm 0.07$ & 14.72 & $0.33\pm 0.10$ & 3.98 & 14.8 \\
M68   &  7 & $15.27\pm 0.06$ & 15.25 & $0.46\pm 0.11$ & 3.85 & 12.3 \\
M30   &  8 & $14.90\pm 0.05$ & 14.88 & $0.32\pm 0.13$ & 3.85 & 12.3 \\
M13   & 17 & $14.46\pm 0.04$ & 14.44 & $0.51\pm 0.17$ & 4.06 & 12.6 \\
N6752 & 18 & $13.34\pm 0.04$ & 13.32 & $0.43\pm 0.17$ & 4.08 & 12.9 \\
N362  & 13 & $15.00\pm 0.05$ & 14.98 & $0.45\pm 0.13$ & 3.87 &  9.9 \\
M5    & 13 & $14.59\pm 0.05$ & 14.57 & $0.54\pm 0.09$ & 4.03 & 11.2 \\
N288  & 12 & $14.97\pm 0.05$ & 14.95 & $0.45\pm 0.13$ & 4.05 & 11.2 \\
47Tuc &  7 & $13.57\pm 0.09$ & 13.55 & $0.55\pm 0.17$ & 4.20 & 12.5 \\
\tableline
\end{tabular}
\label{t:tab4}
\end{table}

\normalsize
\noindent
\begin{table}[ht]
\caption{Systematic effects and total error budget associated with the MS 
fitting distances to GCs} 
\begin{tabular}{lr}
\tableline
\tableline
Effect & $\Delta (m-M)$  \\
\tableline
 Malmquist bias  &   negligible\\
 Lutz-Kelker     &  $\pm 0.02$\\
 Metallicity bias&  only {\it a posteriori} sample\\
 Binaries (in the field) & $\pm 0.02$\\
 Binaries (in clusters)  & $\pm 0.03$\\
 Non solar abundance ratios  & negligible\\
\medskip\\
 Photometric calibrations    & $\pm 0.04$\\
 Reddening scale  & $\pm 0.07$\\
 Metallicity scale& $\pm 0.08$\\
&\\
Total uncertainty ($1\sigma$)& $\pm 0.12$ \\
\tableline
\end{tabular}
\label{t:tab5}
\end{table}

\noindent
\begin{table}[ht]
\caption{Sources of errors in Cluster age determination}
\begin{tabular}{lccc}
\hline \hline
Error source     & Distribution & $\sigma$ & Limits     \\
                 &              &  (Gyr)   &  (Gyr)     \\
\hline
Internal         & gaussian     &   0.5    &            \\
Lutz-Kelker      & gaussian     &   0.25   &            \\
Binaries         & gaussian     &   0.25   &            \\
Metallicity      & flat         &          & $-$1,+1    \\
$[\alpha/{\rm Fe}]$ & flat      &          & $-$0.3,0.8 \\
Reddening        & gaussian     &   0.9    &            \\
Color calibration& gaussian     &   0.5    &            \\
Convection       & flat         &          & $-$0.4,0.4 \\
Code             & flat         &          & $-$0.4,0.4 \\
Diffusion        & flat         &          & $-$1.0,0   \\
Solar $M_v$      & flat         &          & $-$0.3,0.3 \\
\hline
\end{tabular}
\label{t:tab6}
\end{table}

\noindent
\begin{table}[ht]
\caption{True distance modulus to the LMC according to various methods}
\begin{tabular}{lc}
\hline \hline
\multicolumn{2}{c}{Population I}\\
\hline
Cepheids: Trig. parallaxes  & $18.70\pm 0.16$\\
Cepheids: MS fitting        & $18.55\pm 0.08$\\
Cepheids: Baade-Wesselink   & $18.55\pm 0.10$\\
Eclipsing Binaries          & $18.30\pm 0.07$\\
Red clump                   &         ?      \\
Miras                       & $18.54\pm 0.18$\\
SN1987a                     & $18.58\pm 0.05$\\
\hline
\multicolumn{2}{c}{Population II}\\
\hline
Subdwarf fitting            & $18.64\pm 0.12$\\
HB Trig. parallaxes         & $18.50\pm 0.11$\\
RR Lyr: Stat. parallaxes    & $18.38\pm 0.12$\\ 
RR Lyr: Baade-Wesselink     & $18.40\pm 0.2~$\\
RR Lyr: Double mode         & $18.47\pm 0.19$\\
GC Dynamical models         & $18.50\pm 0.11$\\
WD cooling sequence         & $18.4~\pm 0.15$\\
\hline
\end{tabular}
\label{t:tab7}
\end{table}

\begin{figure}
\centerline{\hbox{\psfig{figure=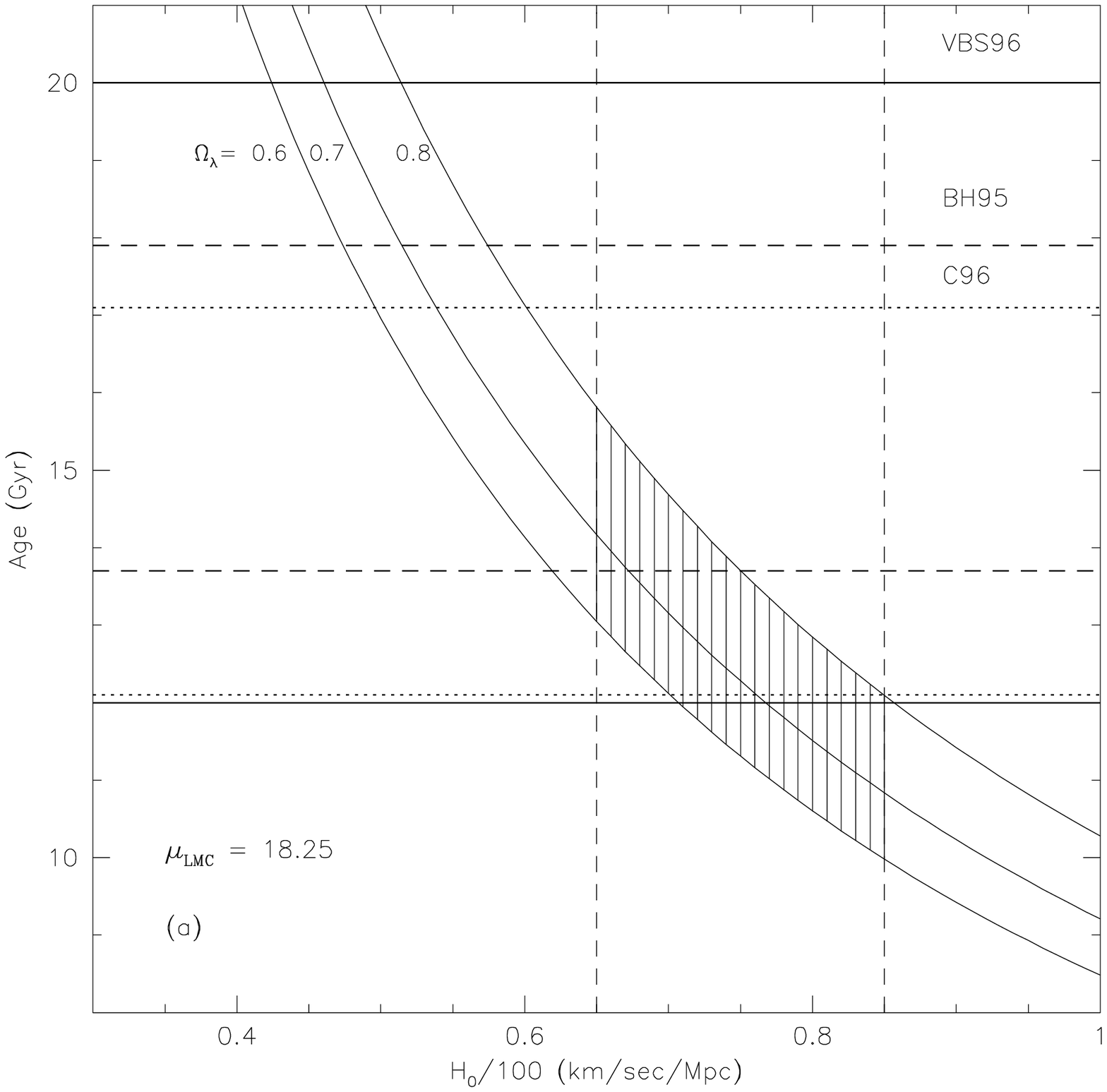,width=13.0cm,clip=}}}
\medskip
\label{f:figure1a}
\end{figure}

\begin{figure}
\centerline{\hbox{\psfig{figure=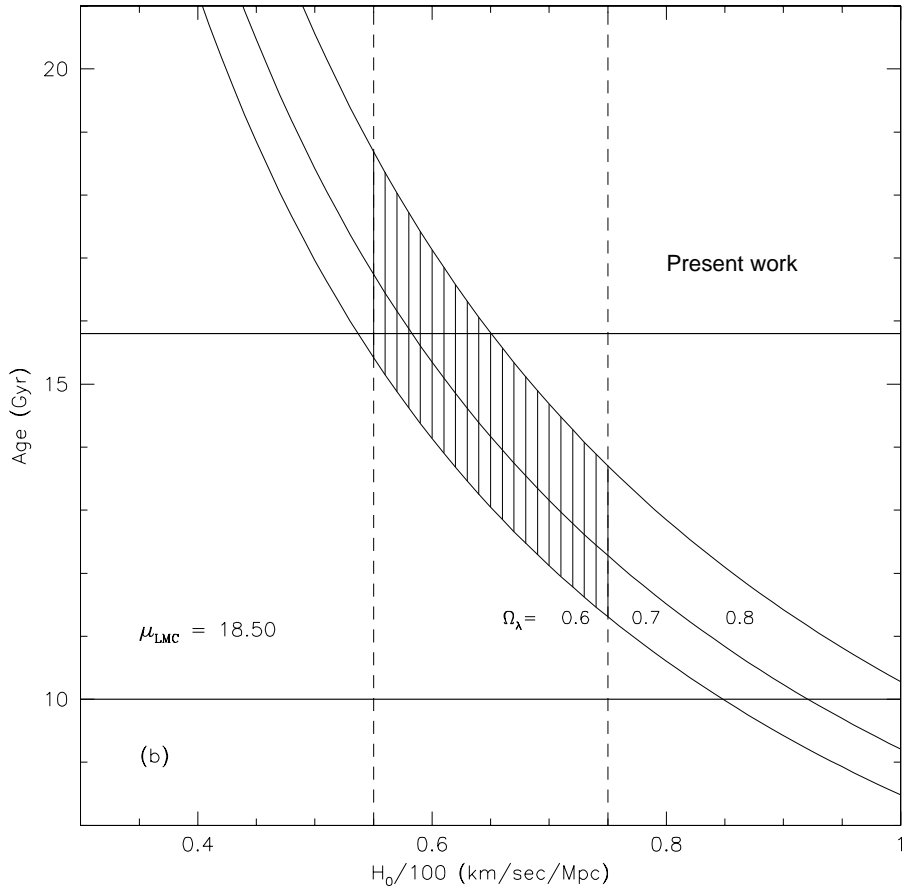,width=13.0cm,clip=}}}
\medskip
\caption{ Age (t$_0$) - H$_0$ relationships, adapted from Turner (1997),
 for various cosmological models of
flat-universe and for different values of $\Omega_\Lambda = 1 - \Omega_m$, 
with $\Omega_m$ in the range suggested by recent type Ia SNe data 
($\Omega_m \sim 0.3 \pm 0.1$). The shaded area is the permitted region 
according to values of H$_0$ consistent with the distance moduli used to 
derive ages for the globular clusters, before : panel {\it (a)} 
(Bolte \& Hogan, 1995, t=$15.8\pm 2.1$~Gyr; 
Chaboyer et al., 1996, t=$14.6\pm 2.5$~Gyr; and VandenBerg, Bolte \&
Stetson, 1996, t=$15^{+5}_{-3}$~Gyr.) and after : panel {\it (b)} the release of 
Hipparcos parallaxes.} 
\label{f:figure1b}
\end{figure}

\begin{figure}
\centerline{\hbox{\psfig{figure=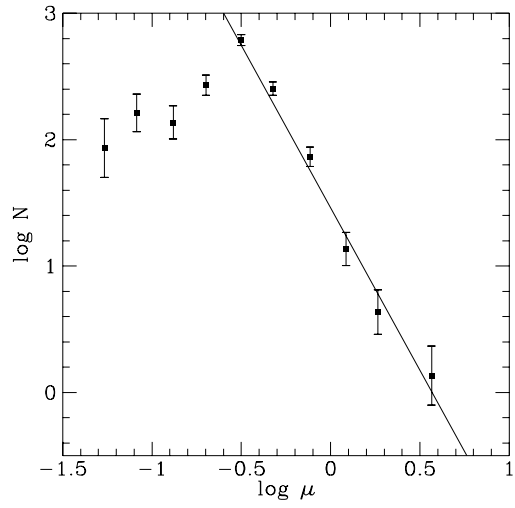,width=13.0cm,clip=}}}
\medskip
\caption{ Proper motion distribution for stars in the present {\it a priori}
sample.}
\label{f:figure2}
\end{figure}

\begin{figure}
\centerline{\hbox{\psfig{figure=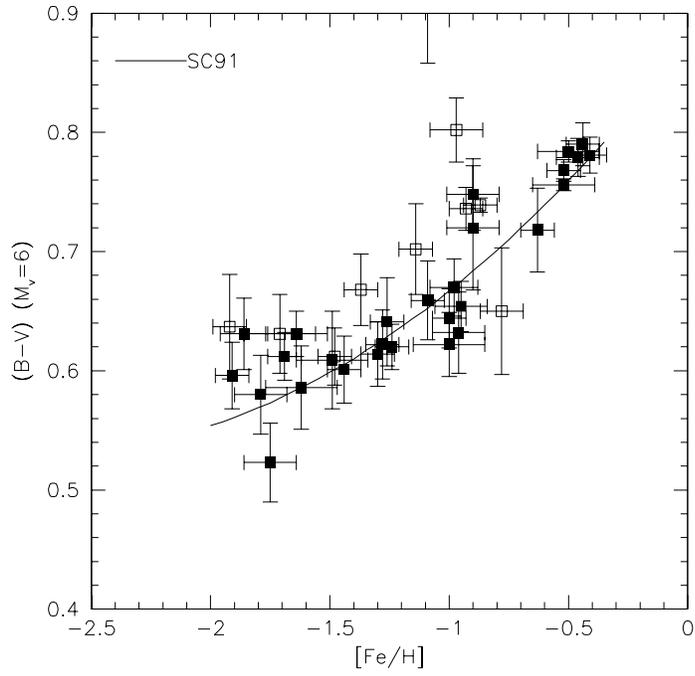,width=13.0cm,clip=}}}
\medskip
\caption{The $(B-V)_{Mv=6}-$[Fe/H] relation for stars in the {\it a priori}
sample with $\Delta\pi/\pi<0.12$. Filled symbols are {\it bona fide} single 
stars; open
symbols are known or suspected binaries. Solid lines are
the theoretical relations by Straniero \& Chieffi (1991).}
\label{f:figure3}
\end{figure}

\begin{figure}
\centerline{\hbox{\psfig{figure=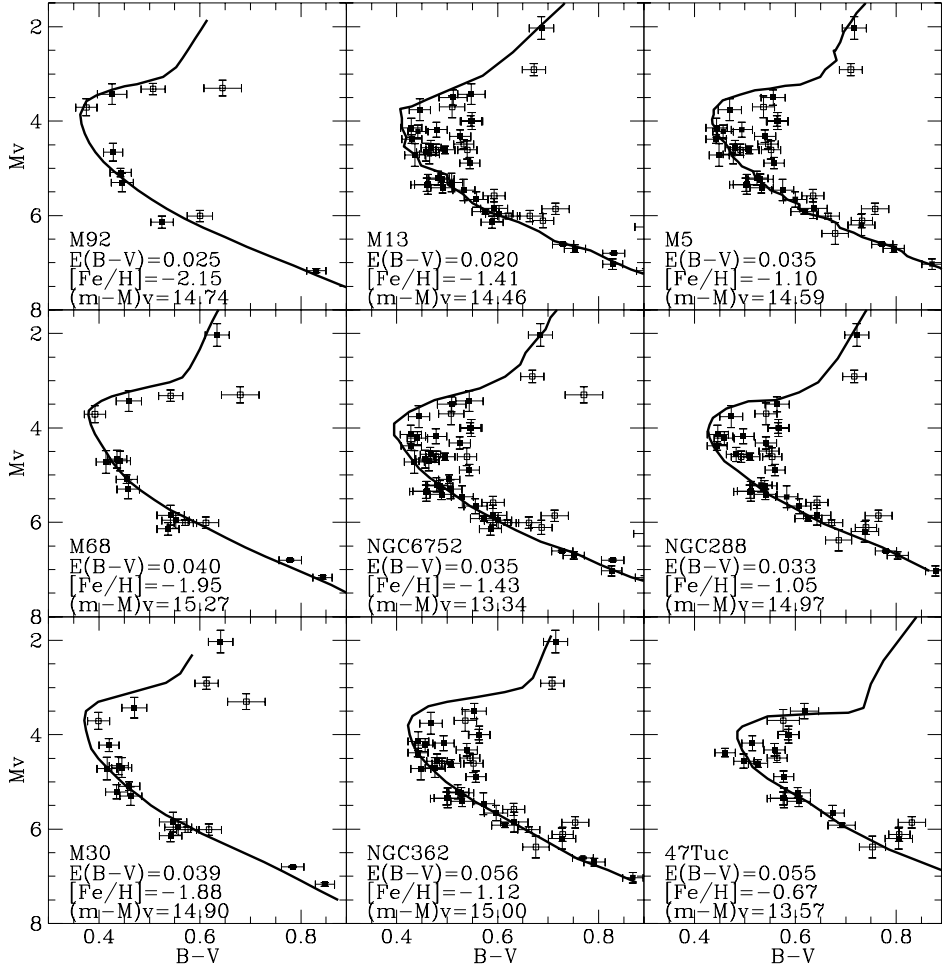,width=13.0cm,clip=}}}
\medskip
\caption{Fits of the fiducial mean loci of the Globular Clusters considered in
this paper with the position of the subdwarfs. Only {\it bona fide} single
stars with $5<M_V<8$\ mag are used in the fits (solid squares). The values of 
the
parameters adopted in the present analysis are shown in each panel.}
\label{f:figure4}
\end{figure}

\begin{figure}
\centerline{\hbox{\psfig{figure=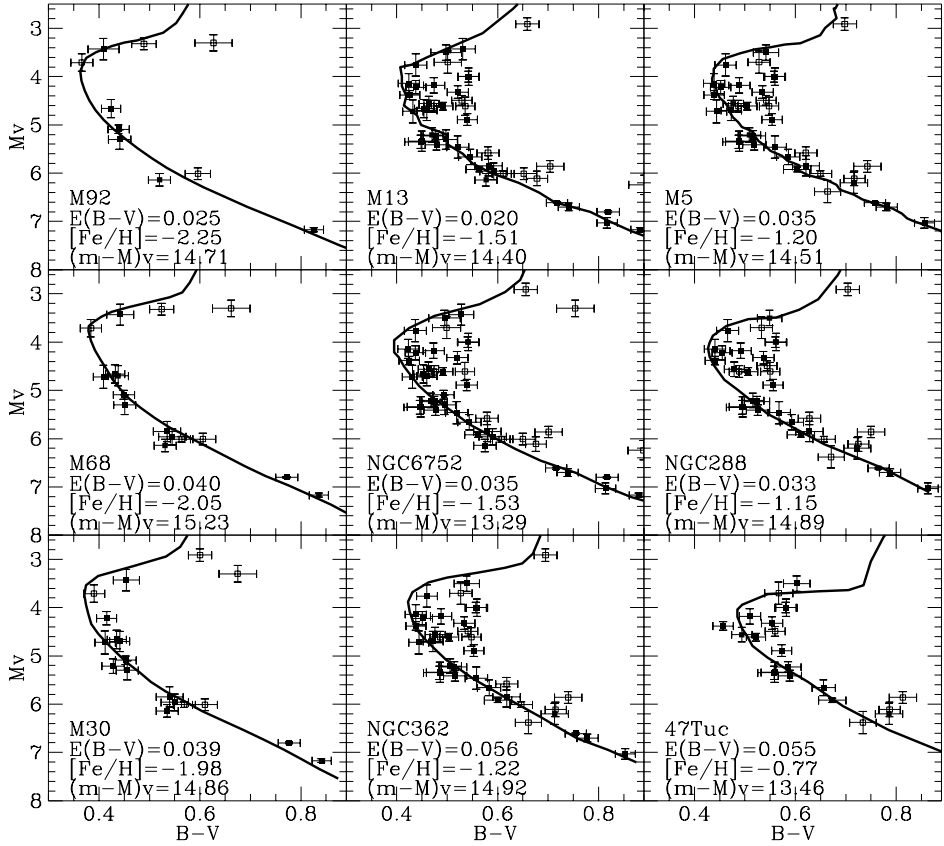,width=13.0cm,clip=}}}
\medskip
\caption{ Same as in Figure 4, but assuming that metallicities of the 
globular clusters from Carretta \& Gratton (1997) are systematically 
underestimate by 0.1 dex.} 
\label{f:figure5}
\end{figure}

\begin{figure}
\centerline{\hbox{\psfig{figure=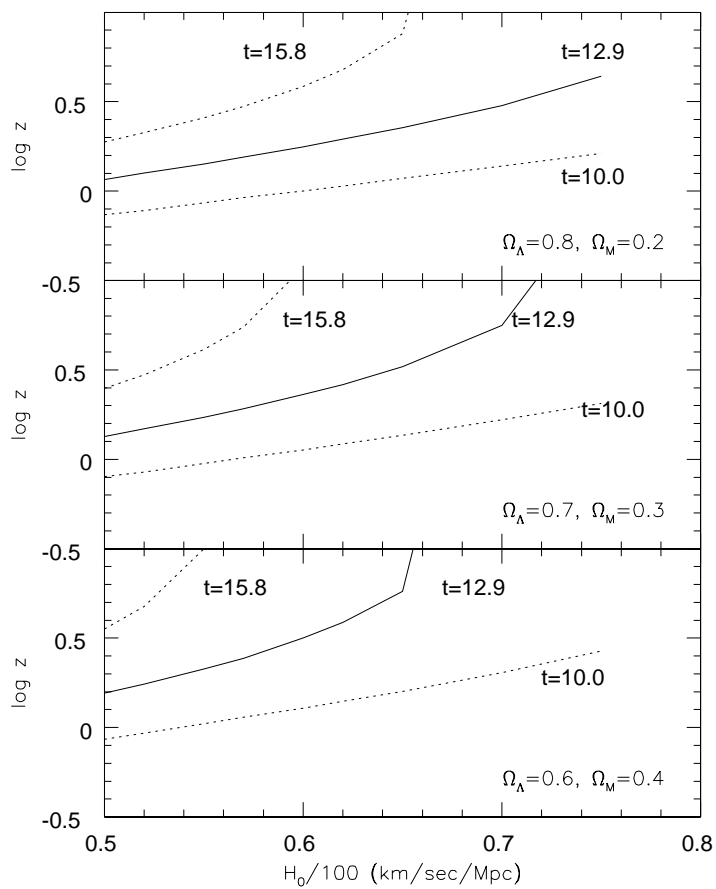,width=13.0cm,clip=}}}
\vspace{1.6cm}
\caption{\small Epoch of formation of the galactic globular clusters against 
various possible values for the Hubble constant in a flat Universe 
$\Omega_\Lambda = 1 - \Omega_m$. The three panels correspond to 3 different
values for $\Omega_m$ (0.2, 0.3 and 0.4). The solid lines shows the favorite 
age of the globular clusters according to this paper (13.2 Gyr), the dashed 
lines represent the 95 $\%$ confidence level.}

\label{f:figure6}
\end{figure}

\end{document}